\documentclass[aps,prb,twocolumn,superscriptaddress,floatfix,10pt,letterpaper,showpacs]{revtex4-1}

\pdfoutput=1

\begin{document}

\begin{titlepage}

\title{Symmetry protected topological phases from decorated domain walls}
\author{Xie Chen}
\affiliation{Department of Physics, University of California, Berkeley, CA 94720, USA.}
\author{Yuan-Ming Lu}
\author{Ashvin Vishwanath}

\affiliation{Department of Physics, University of California, Berkeley, CA 94720, USA.}

\affiliation{Materials Science Division, Lawrence Berkeley National Laboratories, Berkeley, CA 94720, USA.}

\begin{abstract}

Symmetry protected topological (SPT) phases with unusual edge excitations can emerge in strongly interacting bosonic systems and are classified in terms of the cohomology of their symmetry groups. Here we provide a physical picture that leads to an intuitive understanding and wavefunctions for several SPT phases in $d=1,\,2, \,3$ dimensions.
We consider symmetries which include a $Z_2$ subgroup, that allows us to define domain walls. While the usual disordered phase is obtained by proliferating domain walls, we show that SPT phases are realized when these proliferated domain walls are `decorated', i.e. are themselves SPT phases in one lower dimension. 
For example a $d=2$ SPT phase with $Z_2$ and time reversal  symmetry is realized when the domain walls that proliferate are themselves in a $d=1$ Haldane/AKLT state. Similarly, $d=3$ SPT phases with $Z_2\times Z_2$ symmetry emerges when domain walls in a $d=2$ SPT with $Z_2$ symmetry are proliferated. The resulting ground states are shown to be equivalent to that obtained from group cohomology and field theoretical techniques. The result of gauging the Z$_2$ symmetry in these phases is also discussed. An extension of this construction where time reversal plays the role of $Z_2$ symmetry allows for a discussion of several $d=3$ SPT phases. This construction also leads to a new perspective on some well known $d=1$ SPT phases, from which exactly soluble parent Hamiltonians may be derived.  
\end{abstract}

\pacs{71.27.+a, 02.40.Re}

\maketitle

\end{titlepage}



\section{Introduction}
\label{intro}


Symmetry protected topological (SPT) phases are gapped quantum phases with topological properties protectected by  symmetry\cite{Chen2011}. The ground states of SPT phases contain only short-range entanglement and can be smoothly deformed into a totally trivial product state if the symmetry requirement is not enforced in the system. However, with symmetry, the nontrivial SPT order is manifested in the existence of gapless edge states on the boundary of the system which cannot be removed as long as symmetry is not broken. Many SPT phases have been discovered over the past decades\cite{Gu2009,Chen2011b,Gu2012,Levin2012,Lu2012a,Levin2012a,Senthil2012,Vishwanath2012,Lu2012b,Grover2012,Xu2012,Liu2012,Chen2012,Lu2012c,Oon2012,Ye2012,Xu2013,Geraedts2013,Metlitski2013,Wang2013,Burnell2013} and a general structure of the theory of SPT phases is emerging. In 1d, SPT phases have been completely classified for general interacting bosonic/fermionic systems which carry nontrivial projective representations of the symmetry in their degenerate edge states\cite{Chen2011a,Fidkowski2011,Turner2011}. In fact, the Haldane/AKLT phase of spin-one quantum antiferromagnet, is a physical realization of a 1D SPT phase, which is protected by spin rotation or time reversal symmetries. Two dimensional SPT phases were first discovered in topological insulators and superconductors which has subsequently been generalized to three dimensions\cite{Hasan2010,Qi2011,Hasan2011}. Such free fermion SPT phases have realized experimentally\cite{Konig2007,Hsieh2008,Hsieh2009,Hsieh2009a} and also classified completely\cite{Schnyder2008,Kitaev2009}. Recently, it was realized that two and higher dimensional SPT phases exist not only in free fermion systems but also in strongly interacting boson systems and a systematic construction is given based on the group cohomology of the symmetry\cite{Chen2011}. While the construction provides a fixed point description in the bulk, it is hard to access edge dynamics and hence how the system responds to physical perturbations. Also it is not clear in what physically realistic systems can these bosonic SPT phases be realized.


Much progress has been achieved recently in understanding the low energy physics and finding physical realizations for some of the bosonic SPT phases. In 2D, a general understanding of SPT phases with Abelian (and time reversal) symmetry has been given in terms of  Chern-Simons K-matrix theory\cite{Lu2012a}, which also provides a field theory of the protected edge states. A physical realization of the $U(1)$ SPT phases (which are expected to have even integer quantized quantum Hall conductance\cite{Lu2012a}) has been proposed in bosonic cold atom systems with artificial gauge fields \cite{Senthil2012}. The 2D SPT phases with nonabelian $SO(3)$ and $SU(2)$ symmetry was studied with nonlinear sigma model with quantized topological $\theta$ terms and are found to have quantized spin transport\cite{Liu2012}. More recently, some 3D SPT phases have been understood within a field theoretic approach, which predicts surface vortices with projective representations and quantized magnetoelectric responses\cite{Vishwanath2012}. Recently,  3D SPT phases have been discussed from number of different theoretical perspectives , including twisted vortex condensates \cite{Xu2013} and the statistical magneto-electric effect \cite{Metlitski2013}. A useful perspective on SPT phases appears on gauging the symmetry  discussed by Levin and Gu\cite{Levin2012} in 2D, and recently extended in \onlinecite{Wang2013,Metlitski2013,Hung2012,Hung2013,Wen2013} to 3D phases. Finally, we note a special feature of 3D SPT phases is that their surface states could be gapped and fully symmetric if they develop topological order just at the surface\cite{Vishwanath2012,Wang2013,Metlitski2013,Burnell2013}. 



\begin{figure}[htbp]
\begin{center}
\includegraphics[scale=0.25]{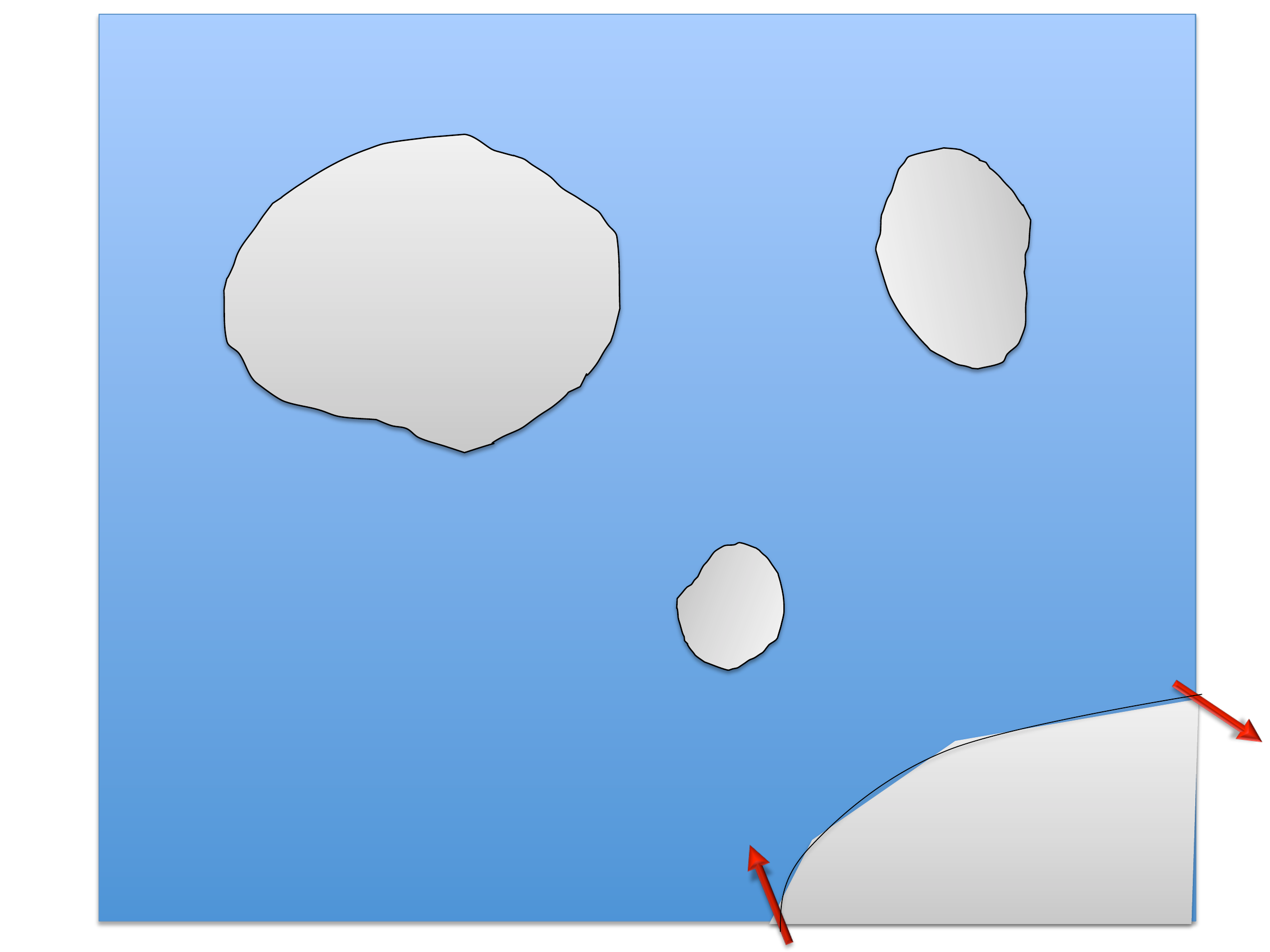}
\end{center}
\caption{SPT phase in $d=2$ with $Z_2$ and time reversal symmetry ($Z_2\times Z_2^T$). A snapshot of the ground state wave function, where blue and grey are oppositely directed domains of the $Z_2$ symmetry. The ground state preserves the $Z_2$ symmetry since it is a superposition of domain configurations. The domain walls themselves (black lines) are in a $d=1$ SPT phase (the Haldane/AKLT phase) protected by time reversal symmetry. When they end at the edge of the system they create Kramers doublets, leading to a gapless edge state.   
}
\label{Fig2D}
\end{figure}

In this paper, we focus on SPT phases with symmetry group $Z_2 \times G$ and present a simple construction of $d$ dimensions SPT phases by decorating the domain walls of $Z_2$ configurations in the bulk with $d-1$ dimensional SPT states with $G$ symmetry. Such a construction naturally reveals some of the special topological features of such phases. For example, it is easy to see that when a domain wall is cut open at the boundary of the system, the end points/loops of the domain wall carry gapless edge states of the $d-1$ dimensional SPT state with $G$ symmetry. The same is true on the flux point/loops in the system when the $Z_2$ part of the symmetry is gauged. In particular, we are going to present the construction of a 2D SPT phase with $Z_2 \times Z_2^T$ symmetry by decorating the 1D $Z_2$ domain walls with Haldane chains and a 3D SPT phase with $Z_2 \times Z_2$ symmetry by decorating the 2D $Z_2$ domain walls with the nontrivial 2D SPT phase with $Z_2$ symmetry. These constructions hence provide a simple understanding of some higher dimensional SPT phases in terms of lower dimensional ones (which are usually better understood) and belongs to a hierarchical construction of SPT phases related to a hierarchy structure of group cohomology. These results are summarized in the figures \ref{Fig2D}, \ref{Fig3D}, \ref{Fig1D}. Besides the $Z_2\times G$ symmetry, similar domain wall picture also allows us to construct SPT phases in 3D with $Z_2^T$, $Z_2^T\times U(1)$ and $Z_2^T\times SO(3)$ symmetry, which produced a $Z_2$ class in each case. 


The paper is organized as follows: In section \ref{2D} we present the construction for the 2D SPT phase with $Z_2\times Z_2^T$ symmetry while section \ref{3D} is devoted to the 3D SPT model with $Z_2 \times Z_2$ symmetry. We describe in detail for both ground state wavefunctions on a lattice, the symmetry action on the edge (and its relation to group cohomology), and an low energy effective theory description of both the bulk and the edge which captures the special domain wall feature of the state. In section \ref{Sec:gauging} we discuss the consequence of gauging the $Z_2$ symmetry.  In section \ref{1Dz2z2} and \ref{3DT} we construct 1D SPT phase with $Z_2\times Z_2$ symmetry and some 3D SPT phases with time reversal symmetry respectively. The relation between the domain wall construction and the K\"{u}nneth formula for group cohomology is discussed in section \ref{Kunneth}. We discuss the implications and the generalizations of our results in section \ref{discussion}. Brief reviews of the group cohomoloy description of SPT phases, the field theory description of 2D and 3D SPT phases  are given in the appendices.



\section{2D SPT phase with $Z_2\times Z_2^T$ symmetry}
\label{2D}



Let's start with the simplest example in this construction: a 2D SPT phase with $Z_2 \times Z_2^T$ symmetry where $Z_2^T$ represents time reversal symmetry. Intuitively, the state is constructed by attaching 1D Haldane chains with time reversal symmetry to the domain walls between $Z_2$ configurations on the plane and then allow all kinds of fluctuations in the domain wall configurations. With such a structure, it is then easy to see that when the system has a boundary where domain walls can end, the end point would carry a spin $1/2$ degree of freedom (edge state of Haldane chain) which transform projectively under time reversal.

In section \ref{2D_H}, we describe a 2D lattice wave function of the  ground state which is gapped and does not break any symmetry. Section \ref{2D_edge} then discusses what happens on the boundary of the system. By identifying the relation between symmetry action on the edge and the nontrivial cocycle of $Z_2\times Z_2^T$ symmetry group, we establishes the existence of nontrivial SPT order in this model. It is known that this SPT phase with $Z_2\times Z_2^T$ symmetry can be described with $U(1)\times U(1)$ Chern-Simons theory with a nontrivial symmetry action. We review this description in section \ref{2D_FT} and demonstrate the nontrivial topological feature of domain walls in the field theory language.


\subsection{Bulk Wave function on 2D lattice}
\label{2D_H}

\begin{figure}[htbp]
\begin{center}
\includegraphics[scale=0.4]{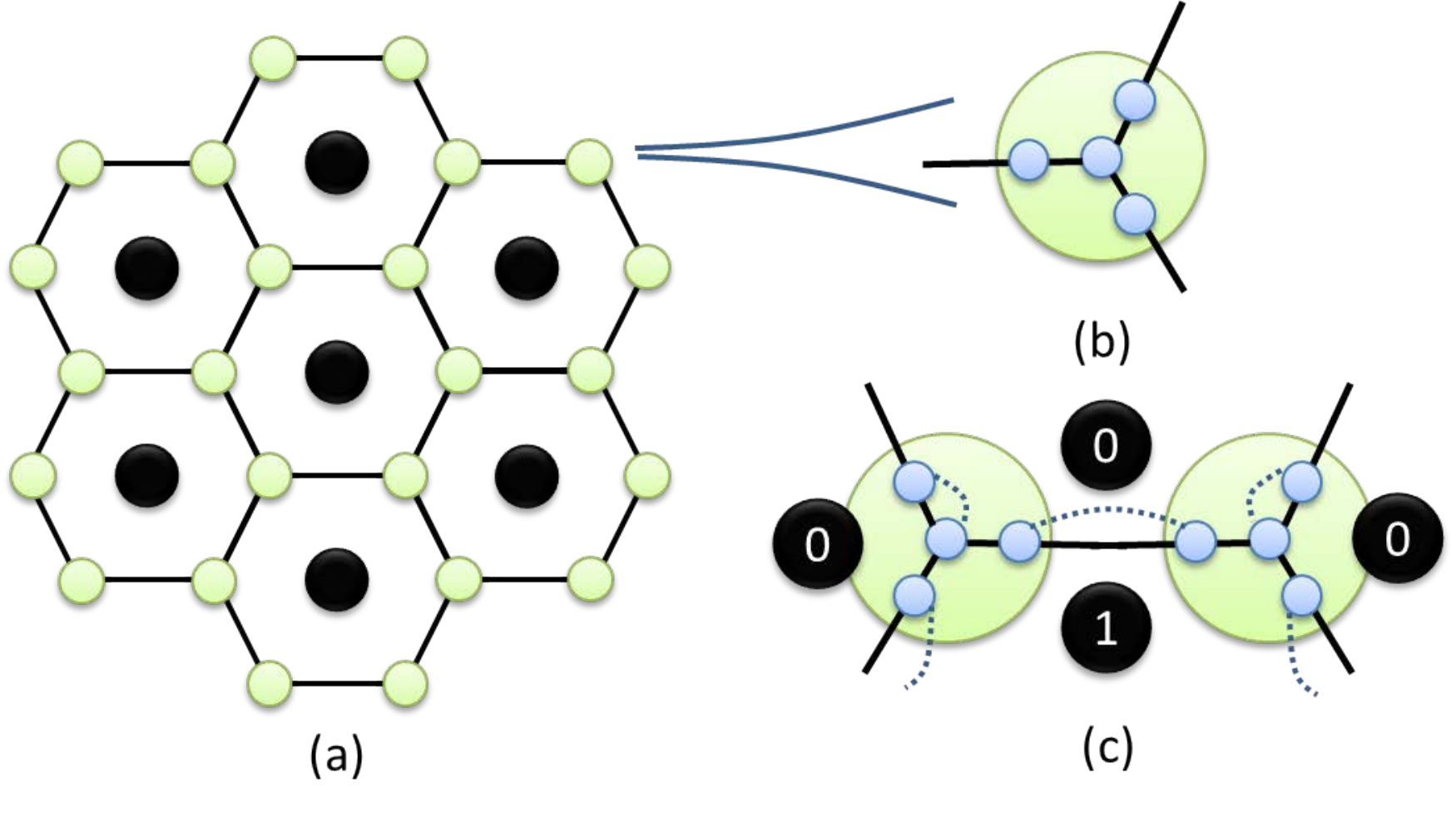}
\end{center}
\caption{Two-dimensional SPT model with $Z_2 \times Z_2^T$ symmetry. (a) each plaquette hosts a $Z_2$ variable (big black dot) (b) each vertex hosts four spin $1/2$'s (small blue dots) (c) two spin $1/2$'s on the same link form a singlet if they are on a $Z_2$ domain wall, other spin $1/2$'s form singlets within each vertex.
}
\label{Z2_T_c}
\end{figure}

Consider a honeycomb lattice as show in Fig. \ref{Z2_T_c} (a) where each plaquette hosts a $Z_2$ variable (big black dot) in state $\ket{0}$ or $\ket{1}$. The $Z_2$ part of the symmetry flips $\ket{0}$ into $\ket{1}$ and $\ket{1}$ into $\ket{0}$. At each vertex, there are four spin $1/2$'s (one at the center and three on links as shown in Fig. \ref{Z2_T_c} (b)). On each spin $1/2$ time reversal symmetry acts as $\cT=i\si_yK$ and satisfies $\cT^2=-1$. On the total Hilbert space of each vertex, time reversal still satisfies $\cT^2=1$ and forms a linear representation of $Z_2^T$. Now consider a Hamiltonian term $V$ which enforces that in the ground state two spin $1/2$'s on the same link form a singlet pair if the link is on a $Z_2$ domain wall while the remaining spin $1/2$'s which are not on a domain wall form singlets within each vertex (there are always an even number of these at each vertex). The effect of this term can be thought of as attaching Haldane chains to all the $Z_2$ domain walls. A possible configuration in the ground state is shown in Fig. \ref{Z2_T_c} (c) where the dotted lines represent singlet pairing. A tunneling term between the $Z_2$ configurations $\sum_i \tau_x^i$ is then added to the Hamiltonian which in the low energy sector of $V$ flips $Z_2$ variables together with the related singlet configurations. Therefore, the ground state of the total Hamiltonian
\be
H=uV+\sum_i \tau_x^i
\ee
when $u$ is very large is an equal weight superposition of all possible consistent configurations of $Z_2$ variables and singlets. The ground state is unique, gapped and preserves $Z_2\times Z_2^T$ symmetry.


\subsection{Edge state}
\label{2D_edge}


\begin{figure}[htbp]
\begin{center}
\includegraphics[scale=0.4]{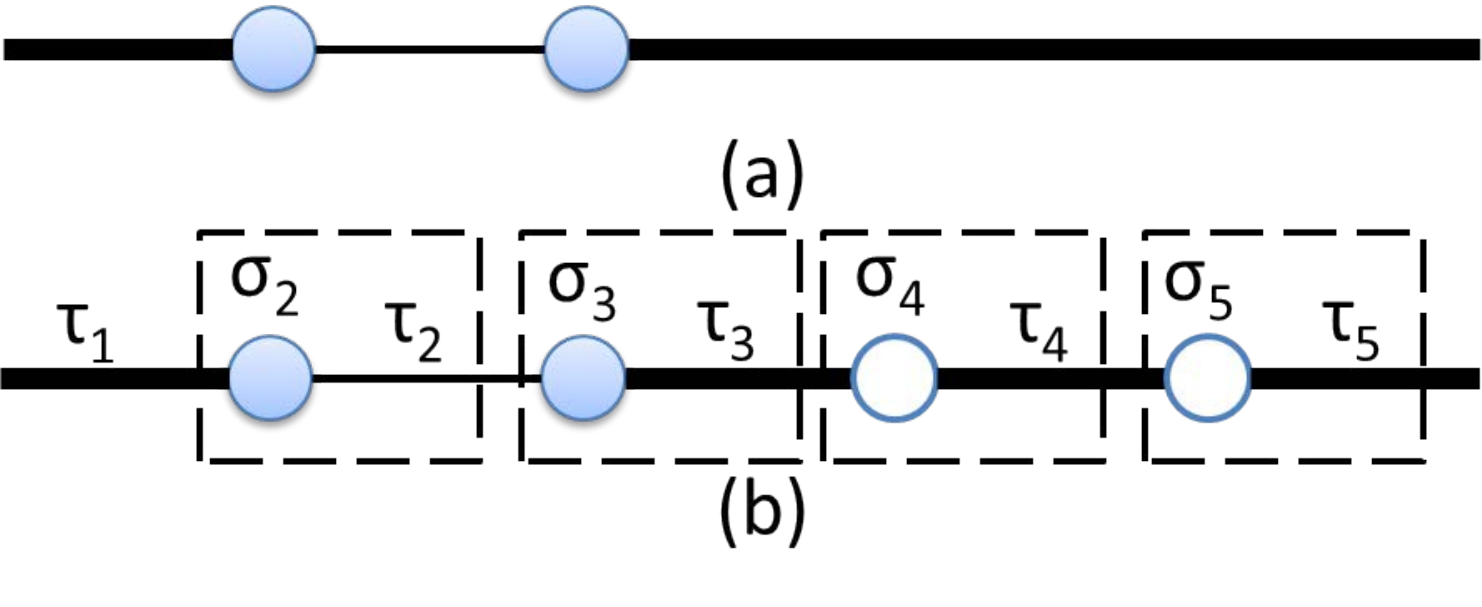}
\end{center}
\caption{Two equivalent descriptions of the boundary of 2D SPT model with $Z_2\times Z_2^T$ symmetry. (a) Thick bonds represent $Z_2$ variable in state $\ket{1}$ and thin bonds $\ket{0}$, spin $1/2$ exists on their domain wall. (b) One $Z_2$ variable per bond and one spin variable per vertex. Time reversal acts on spins in a way dependent of neighboring $Z_2$ configurations. Dotted boxes represent local degrees of freedom on the 1D boundary labeled by group elements.
}
\label{Z2_T_b}
\end{figure}

Interesting things happen when the system has a boundary. When the system is cut open, the $Z_2$ domain walls will have end points on the boundary. Because the spin singlets are tied to domain walls in the bulk, on the boundary there will be isolated spin $1/2$'s on the end points of $Z_2$ domain walls. Imagine that we cut the system through the middle of plaquettes. (In order not to cut through degrees of freedom, we can first split the $Z_2$ variable into two, one on each side of the cut and add a Hamiltonian term for them to be equal. This does not violate the symmetry of the system and is therefore allowed.) The 1D boundary degrees of freedom then contain two parts: the bond $Z_2$ variables which come from plaquettes on the cut and the vertex spin variables when the neighboring bonds contain different $Z_2$ configurations, as shown in Fig. \ref{Z2_T_b} (a). Symmetry acts by flipping the $Z_2$ variable and time reversing the spins.

To study the boundary as an effective one dimensional system, we will use a slightly different description of the boundary degrees of freedom. In the previous description, the existence of vertex degrees of freedom are dependent on the bond degrees of freedom. Equivalently, we can also think of the boundary as a local 1D system with independent degrees of freedom on the bonds and the vertices, as shown in Fig. \ref{Z2_T_b} (b). That is, we can imagine that there exist pseudo-spins (empty circles) on the vertices when neighboring $Z_2$ variables are the same. On the pseudo-spins time reversal acts simply as $\cT=\si_xK$ and satisfies $\cT^2=1$. The total symmetry action is then
\be
\begin{array}{ll}
Z_2: & \tau_x \text{\ \ on each $\tau$ spin}\\ \nonumber
Z_2^T: & i\sigma_y K \text{\ \ if $\sigma$ is on $\tau$ domain wall} \\ \nonumber
   & \sigma_x K \text{otherwise}
\end{array}
\ee

It is easy to see that such a description of the boundary can be reduced to the previous one by polarizing each pseudo-spin in the $z$ direction. Because $\cT^2=1$ on each pseudo-spin, polarizing them does not break time reversal symmetry. Using this description, we can see explicitly how this model is related to the nontrivial 3rd cocycle of the symmetry group and therefore contains nontrivial SPT order. We can also write down effective Hamiltonians for the boundary and solve for the low energy dynamics, as we demonstrate in the following.

In terms of the local degrees of freedom on the 1D edge state, $\si$ and $\tau$, the $Z_2$ symmetry action on the edge reads
\be
Z_2: \prod_i \tau_x^i
\label{sym2_Z2}
\ee
and time reversal acts as
\be
Z_2^T: \prod_{i,i+1} \Big(\frac{I+\tau_z^i\tau_z^{i+1}}{2}\si^{i+1}_x+\frac{I-\tau_z^i\tau_z^{i+1}}{2}i\si^{i+1}_y\Big)K
\label{sym2_T}
\ee
We can write down an effective Hamiltonian satisfying the symmetry and solve for the dynamics on the 1D edge. Some simple interaction terms which satisfy both the $Z_2$ symmetry (Eq.\ref{sym2_Z2}) and the time reversal symmetry (Eq.\ref{sym2_T}) include $\tau_x^i+\si_z^{i}\tau_x^i\si_z^{i+1}$ and $\si_x^i+\tau_z^{i-1}\si_x^i\tau_z^{i}$. Therefore, a possible form of the dynamics of the edge is given by Hamiltonian
\be\label{1d edge ham}
H_e=\sum_i \tau_x^i+\si_z^{i}\tau_x^i\si_z^{i+1} + \si_x^i+\tau_z^{i-1}\si_x^i\tau_z^{i}
\ee
This Hamiltonian can be mapped exactly to an $XY$ model by applying unitary transformations $U=\Big[\prod_n\imth\frac{\tau^z_n+\tau^x_n}{\sqrt2}\Big]\cdot\Big[\prod_n(-1)^{(1-\tau_z^n)(1-\si_z^n)/4}\Big]$ to each pair of $\tau_i$ and $\si_i$ variables. The Hamiltonian that we get after these transformations is
\be
\t H_e=UH_eU^{-1}=\sum_i \tau_x^i\si_x^{i+1}+\tau_z^i\si_z^{i+1}+\si_x^i\tau_x^i+\si_z^i\tau_z^i
\label{XZ}
\ee
and the symmetry transformations are mapped to
\be
\t Z_2: \prod_i \tau_z^i\si_z^i
\label{Z2_XZ}
\ee
and
\be
\t Z_2^T: \prod_i \Big(\frac{\tau_x^i+\tau_x^{i-1}}{2}\si_x^i+ \frac{\tau_x^i-\tau_x^{i-1}}{2}i\si_y^i\Big) K
\label{Z2T_XZ}
\ee
In order to see how the symmetry acts on the low energy effective theory, we diagonalize the $XY$ Hamiltonian $\t H_e$, identify the free boson modes in the low energy eigenstates and calculate the action of the symmetry on the low energy states. The low energy states are labeled by quantum numbers $n_k \in Z$ and $\bar n_k \in Z$, $k=0,1,2...$, where $n_0$ and $\bar n_0$ label the total angular momentum and the winding number of the boson field respectively and $n_k$ and $\bar n_k$, $k>0$, label the left/right moving boson modes. The $Z_2$ symmetry acts by mapping state $\ket{n_0,\bar n_0,\{n_k,\bar n_k\}}$ to $(-)^{\sum_{k>0} n_k+\bar n_k}\ket{-n_0,-\bar n_0, \{n_k, \bar n_k\}}$ and time reversal symmetry acts by mapping state $\ket{n_0,\bar n_0,\{n_k,\bar n_k\}}$ to $(-)^{\sum_{k\geq 0} n_k+\bar{n}_k}\ket{n_0,-\bar n_0, \{\bar n_k, n_k\}}$, which is consistent with the low energy description given in the next section where $Z_2$ symmetry acts as $\phi_1 \to -\phi_1$, $\phi_2 \to -\phi_2$ and time reversal acts as $\phi_1 \to -\phi_1+\pi$, $\phi_2 \to \phi_2+\pi$ on the chiral boson fields $\{\phi_{1,2}\}$. 

\subsection{Effective field theory description}
\label{2D_FT}
As discussed in \Ref{Lu2012a}, SPT phases in 2D may be characterized by the unusual transformation properties of their edge states, under the action of symmetry. The simplest situation, which captures a large fraction of SPT phases, is a Luttinger liquid edge with a single gapless bosonic mode. The edge fields are characterized by compact conjugate bosonic fields $\phi_1,\,\phi_2$ such that
\be
\left [\phi_1(x),\,\frac{\partial_{x'}\phi_2(x')}{2\pi}\right] = i\delta (x-x')
\label{commutation}
\ee

Physically, $e^{i\phi_1(x)}$ inserts a boson at point $x$ along the edge, while $e^{i\phi_2}$ inserts a phase slip. From the discussion in the previous section and in Appendix \ref{app:2d}, we see that the 2D SPT phase of interest implies the following transformation law:
\begin{eqnarray}
Z_2: \,&& \phi_1 \rightarrow \phi_1+\pi\\ \nonumber
&&\phi_2\rightarrow \phi_2\\ \nonumber
Z_2^T: \,&& \phi_1 \rightarrow -\phi_1\\ \nonumber
&& \phi_2 \rightarrow \phi_2+\pi
\end{eqnarray}

Note, this edge cannot be gapped out without breaking symmetry. Terms that gap the edge such as $\cos \phi_{1,2}$ are forbidden by symmetry. If either one of the symmetries is broken, then a trivial edge is possible. Finally let us verify that a domain wall of the $Z_2$ order, terminating at the edge carries a Kramer's pair, as the pictorial description implies.

The operator that creates a domain wall, rotates $\phi_1$ by $\pi$ everywhere (say) to the right of the point $x$. By the commutation relation \ref{commutation}, this is identified with the operator

\be
D(x) = e^{i\phi_2(x)/2}
\ee
Now, let us consider how the domain wall insertion operator transforms under time reversal $\mathcal T$. In particular we would like to calculate the action of acting twice with time reversal and see if $\mathcal T^2=\pm 1$. Now $D(x)\overset{{\mathcal T}}\longrightarrow e^{-i(\phi_2+\pi)/2}=-iD^*(x) $. Applying this again we find:

\be
D(x) \overset{\mathcal T^2}\longrightarrow=-D(x)
\ee

thus we find that for this operator ${\mathcal T^2} = -1$. Hence it must carry a Kramer's degeneracy as expected from the physical picture.


\subsection{Connection to Group Cohomology}
\label{2D_GC}
Now let us establish the nontrivial SPT order in this state by identifying the connection of the symmetry actions on the boundary with nontrivial third cocycles. When we think of the boundary as a local 1D system with independent bond variables and vertex variables (i.e. treating spins and pseudo-spins as equivalent variables), symmetry no longer acts on the degrees of freedom independently. The $Z_2$ part of the symmetry $Z_2$: $\prod_i \tau_x^i$ still acts on each $Z_2$ variable independently. However, time reversal symmetry on the vertices depends on the $Z_2$ configurations on the bonds, as shown in Eq.\ref{sym2_T}. This is exactly the signature of SPT phases.

Indeed, as was shown in \Ref{Chen2011}, the boundary of $d$D STP phases can be thought of as a $d-1$D local system where the symmetry acts in a non-onsite way. Without symmetry, the boundary can be easily gapped out. If the symmetry acts in an on-site way, the boundary can be simply gapped out by satisfying the symmetry on each site. However, with non-onsite symmetry related to nontrivial group cocycles, the boundary must remain gapless as long as symmetry is not broken.

In particular, in the group cohomology construction as reviewed in appendix \ref{app:GC_construct}, the boundary local degrees of freedom are labeled by group elements $\al_i$ of the symmetry group and the action of the symmetry operator involves two parts:
\begin{itemize}
\item{1. changing the local states $\ket{\al_i}$ to $\ket{\al\al_i}$,}
\item{2. multiplying a phase factor given by nontrivial cocycles to each pair of $\al_i$ and $\al_{i+1}$.}
\end{itemize}
More specifically, on the 1D boundary of 2D SPT phases, the symmetry acts as
\be
O(\al)\ket{\al_1,....,\al_N}=\prod_i f^{\al}(\al_i,\al_{i+1})\ket{\al\al_1,...,\al\al_N}
\label{sym_1d}
\ee
where $f^{\al}(\al_i,\al_{i+1})$ is a phase factor given by the nontrivial 3rd cocycle $\om_3$
\be
f^{\al}(\al_i,\al_{i+1})=\om_3(\al_i^{-1}\al_{i+1},\al_{i+1}^{-1}\al^{-1},\al)
\ee

Now we can show that the symmetry action on the boundary of the $Z_2\times Z_2^T$ model we constructed is of exactly this form. We can consider the pair of $Z_2$ variables $\al_i=(\tau_i,\si_i)$ (dotted box in Fig. \ref{Z2_T_b} (b)) as labeling group elements  in the group $Z_2 \times Z_2^T$. That is, we consider the $Z_2$ state $\ket{0}/\ket{1}$ of $\tau$ as labeling the trivial/nontrivial element in group $Z_2$ and the spin state $\ket{\up}/\ket{\down}$ of $\sigma$ as labeling the trivial/nontrivial element in group $Z_2^T$. Then the $Z_2$ part of the symmetry is the following mapping
\be
\ket{\tau_i} \to \ket{\bar{\tau_i}}
\ee
and the time reversal part of the symmetry (given by either $\cT=\si_xK$ on the pseudo-spin or $\cT=i\si_yK$ on the spin) also involves the mapping
\be
\ket{\si_i} \to \ket{\bar{\si_i}}
\ee
Therefore, a general symmetry operation labeled by $\al=(\tau,\si)$, $\tau\in Z_2$, $\si\in Z_2^T$, will first change group elements labels in each box
\be
\ket{\al_i=(\tau_i,\si_i)} \to \ket{\al\al_i=(\tau\tau_i,\si\si_i)}.
\ee

Moreover, the time reversal symmetry also adds a $(-1)$ phase factor when the neighboring $Z_2$ variables are different and when the vertex variable was originally in state $\ket{\up}$. Such a phase factor on spin $\sigma_i$ can be written as
\be
(-\si_z^{i+1})^{\frac{1-\tau_z^i\tau_z^{i+1}}{2}}
\ee
Here $\si_z=1$ in $\ket{\up}$, $\si_z=-1$ in $\ket{\down}$ and $\tau_z=1$ in $\ket{\up}$, $\tau_z=-1$ in $\ket{\down}$. Therefore, the symmetry action on the boundary can be put exactly into the form of Eq. \ref{sym_1d} with
\be
f^{\al}(\al_i,\al_{i+1}) = (-\si_z^{i+1})^{\frac{1-\tau_z^i\tau_z^{i+1}}{2}},
\ee
when the symmetry action $\al$ involves time reversal and
\be
f^{\al}(\al_i,\al_{i+1}) = 1,
\ee
when $\al$ does not involve time reversal. We can reorganize the variables and write the phase factor as a function of $\al_i^{-1}\al_{i+1}$, $\al_{i+1}^{-1}\al^{-1}$
\be
\om_3(\al_i^{-1}\al_{i+1},\al_{i+1}^{-1}\al^{-1},\al)=f^{\al}(\al_i,\al_{i+1})
\ee
It can be checked that $\om_3(\al_i^{-1}\al_{i+1},\al_{i+1}^{-1}\al^{-1},\al)$ is a nontrivial third cocycle of group $Z_2 \times Z_2^T$, using the cocycle conditions introduced in appendix \ref{GC}. Therefore, we can show using methods in \Ref{Chen2011b} that the boundary must be either gapless or symmetry breaking and the 2D bulk is in a nontrivial SPT phase.





\section{3D SPT phase with $Z_2\times Z_2$ symmetry}
\label{3D}



Now we go one dimension higher and consider the $Z_2 \times Z_2$ group. To differentiate the two $Z_2$'s, we write them as $Z_2$ and $\t Z_2$. We can start with similar constructions in the bulk where the nontrivial $\t Z_2$ SPT phase in two dimension is attached to the domain wall of the $Z_2$ configuration in the cubes. A simple understanding of the ground state wave function exists starting from Levin and Gu's construction of 2D SPT phase with $Z_2$ symmetry. It was shown that in the 2D SPT model the ground state wave function takes the simple form of
\begin{equation}
\psi_{2D} ({\mathcal C}) = (-1)^{N_{\mathcal C}}
\label{wfn}
\end{equation}
where $N_{\mathcal C}$ is the number of domain walls in configuration ${\mathcal C}$. Such a wave function has gapless edge states when the system has a boundary.  Now consider the symmetry $Z_2\times \t Z_2$. This has elements $\{ 1,\, g_1,\,g_2,\,g_3 \}$. We will pick two of the three nontrivial elements say $g_1,\,g_2$. Choose any one of these two generators (say $g_1$) and consider domain walls in 3D between regions that break this symmetry in opposite ways. This defines closed 2D manifolds. Now, on these closed manifolds the domain walls of the second generator ($g_2$) are examined and number of closed loops counted (essentially these loops are intersections of domain walls of $g_1$ and $g_2$). Now, one uses this set of closed loops and defines a wave function as in Eq.\ref{wfn}.

\begin{equation}
\psi_{3D} ({\mathcal C}) = (-1)^{N^{g_1g_2}_{\mathcal C}}
\label{wfn3D}
\end{equation}

\begin{figure}[htbp]
\begin{center}
\includegraphics[scale=0.3]{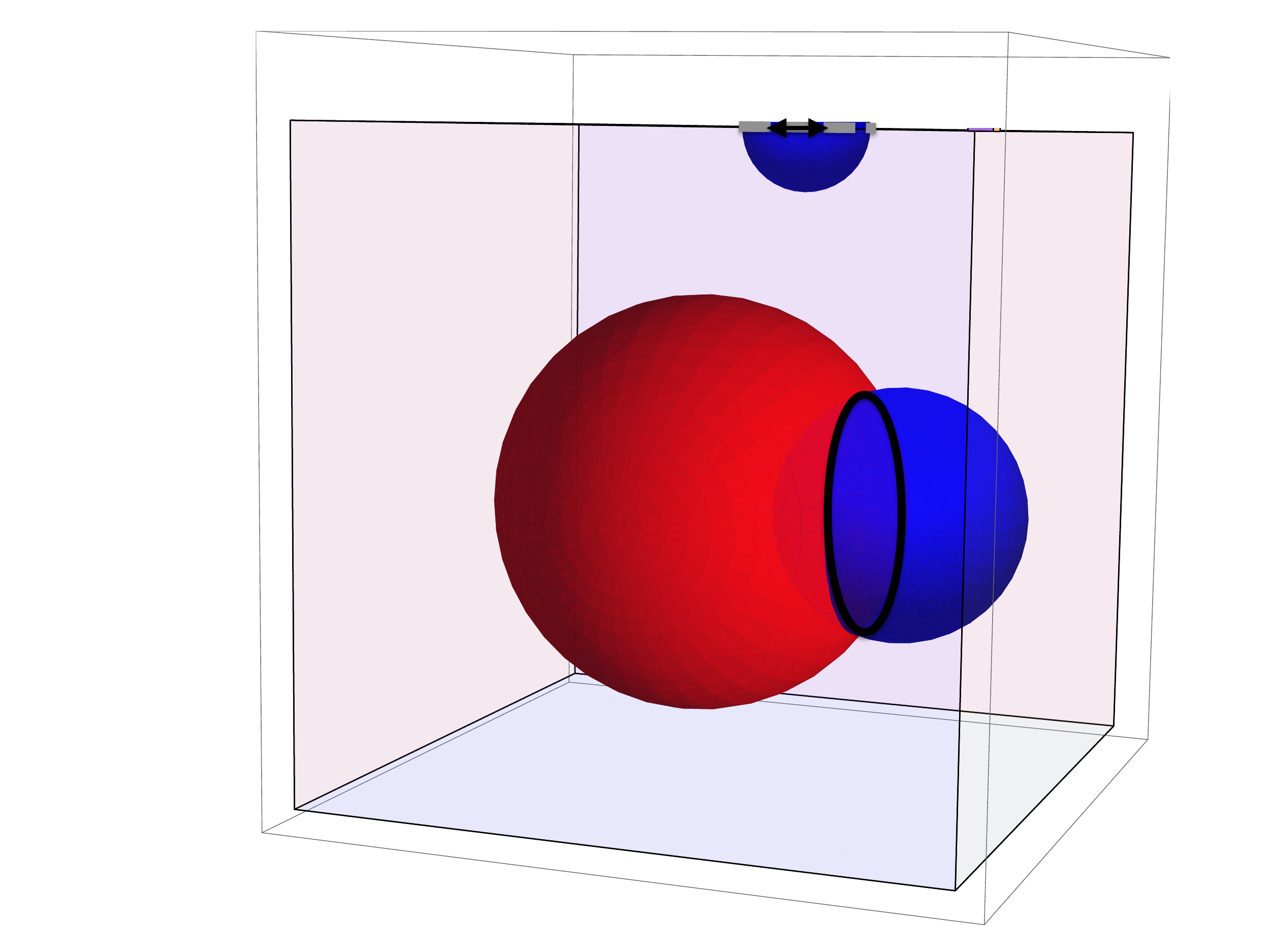}
\end{center}
\caption{SPT phase in $d=3$ with $Z_2\times \t Z_2$ symmetry. Red (blue) surfaces represent domain walls of the $Z_2$ ($\t Z_2$) symmetry. They intersect along curves (black lines). The ground state wave function is a superposition of all domain wall configurations which differ by a sign depending on whether there are an even or odd number of  intersection curves. This automatically implies a protected edge state when a domain wall intersects the surface of the sample (grey dashed line). 
}
\label{Fig3D}
\end{figure}

This is the wave function of a $Z_2\times \t Z_2$ symmetric state where $N^{g_1g_2}_{\mathcal C}$ is the number of closed loops formed by the intersection of the $g_1$ and $g_2$ domain walls in configuration $\mathcal C$ (see Figure \ref{Fig3D}). What is the physical consequence of this wave function? Consider breaking one of the $Z_2$ symmetries and forming a domain wall (of say $g_1$). Now, the domain wall is the edge state of the 2D SPT phase protected by the unbroken $Z_2$ symmetry. This is also evident from the wave function. Due to this nontrivial topological features, the wave function describes a nontrivial SPT phase.

Because there are three different ways to pick two generators out of the three nontrivial elements in $Z_2\times \t Z_2$, we can construct three different SPT wave function in this way. Therefore, this construction allows us to access all possible nontrivial phases classified using group cohomology theory. In the following sections we will present a more detailed study of this construction. Starting from a bulk Hamiltonian in section \ref{3D_H}, we analyze its edge state in section \ref{3D_edge} and demonstrate its relation to nontrivial group cocycles. In section \ref{3D_FT}, we present the field theory description of these phases \cite{Vishwanath2012}.

\subsection{Ground State Wave-function on a 3D lattice}
\label{3D_H}

\begin{figure}[htbp]
\begin{center}
\includegraphics[scale=0.3]{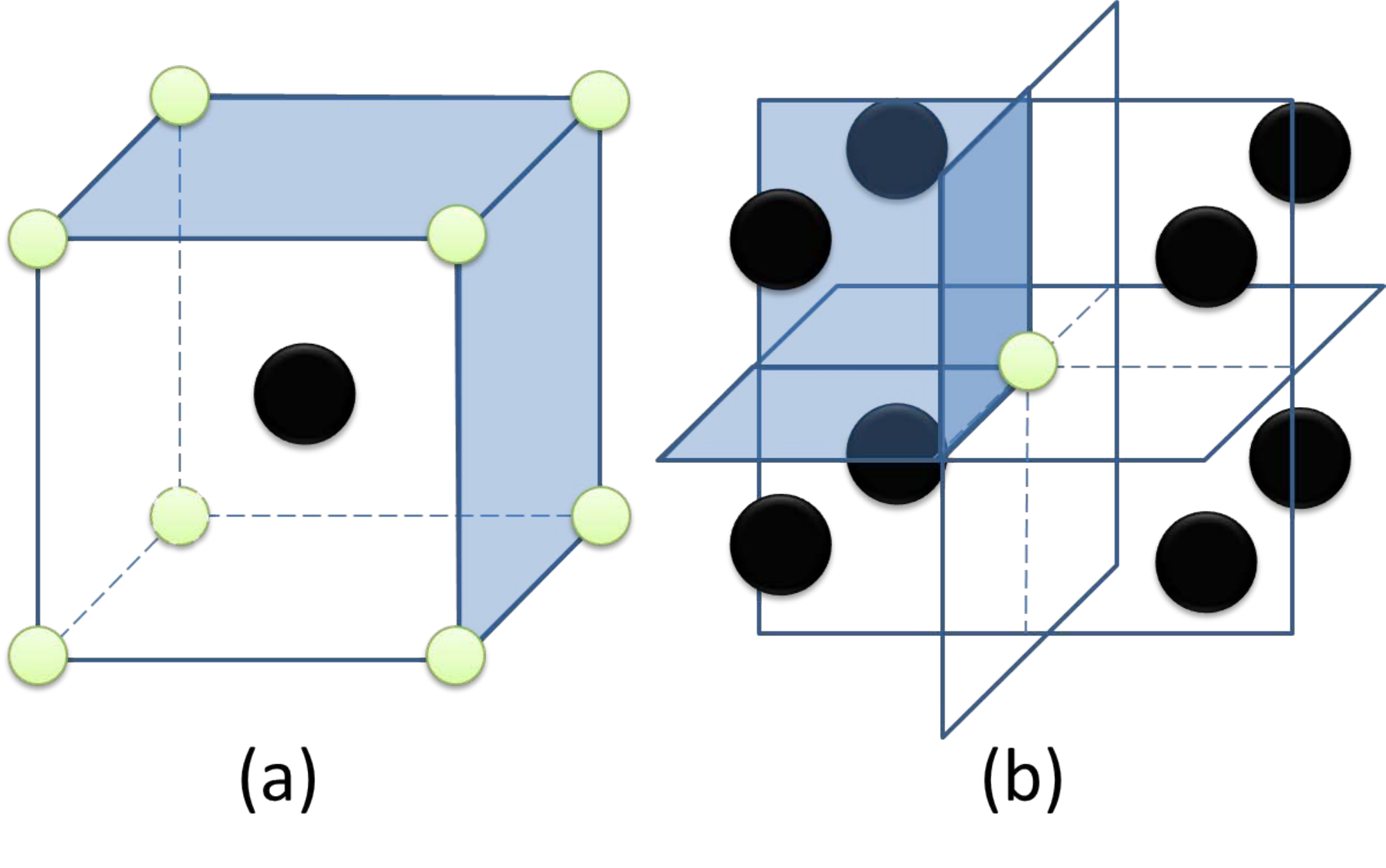}
\end{center}
\caption{Three dimensional SPT model with $Z_2 \times \t Z_2$ symmetry. Each cube hosts a $Z_2$ variable (big black dot) and each vertex hosts a $\t Z_2$ variable (small green dot) (a) a cube and the vertices around it (b) a vertex and the cubes around it. Shaded surfaces represent $Z_2$ domain walls.
}
\label{Z2_Z2_c}
\end{figure}

Consider a 3D cubic lattice where each cube in the bulk hosts a $Z_2$ variable (big black dot in Fig. \ref{Z2_Z2_c}) and the $Z_2$ symmetry flips $\ket{0}$ to $\ket{1}$ and $\ket{1}$ to $\ket{0}$. Each vertex in the 3D bulk hosts a $\t Z_2$ variable (small green dot in Fig. \ref{Z2_Z2_c}) and the $\t Z_2$ symmetry flips $\ket{\t 0}$ to $\ket{\t 1}$ and $\ket{\t 1}$ to $\ket{\t 0}$. To define the Hamiltonian of the system, we first fix a $Z_2$ configuration in the cubes and define the interaction between the $\t Z_2$ vertices. Start with a magnetic field in the $x$ direction on all the $\t Z_2$ spins. If a vertex is on the domain wall between $Z_2$ configurations, modify the Hamiltonian term $\si_x$ by an extra factor of
\be
(-)^{n_{dwp}}
\ee
where $n_{dwp}$ is the number of $\t Z_2$ domain wall pairs along the loop of all nearest neighbor $\t Z_2$ spins on the same $Z_2$ domain wall as the original $\t Z_2$ spin. \cite{Note1} If two or more $Z_2$ domains walls touch at the $\t Z_2$ vertex, then one factor is added for each domain wall. (We can pick a particular way to separate the domain walls at each vertex.) Now it is easy to see that the $\t Z_2$ spins form 2D nontrivial $\t Z_2$ SPT states on the $Z_2$ domain wall. Away from the domain wall, they are polarized in the $x$ direction. Now we include tunneling between the $Z_2$ configurations $\tau_x$, which in the low energy sector of the previous Hamiltonian term not only flips the $Z_2$ spins but also changes the corresponding $\t Z_2$ interaction pattern in its neighborhood. All these terms are local. 
The ground state is then unique and gapped and is the equal weight superposition of all $Z_2$ configurations together with the $\t Z_2$ SPT state on its domain wall and polarized $\t Z_2$ spin away from its domain walls. The ground state is symmetric under the $Z_2 \times \t Z_2$ symmetry.


\subsection{Connecting the edge state to Group Cohomology}
\label{3D_edge}


Imagine that we cut the system open and expose the boundary. Note that when doing the cut, we need to be careful and not to break the symmetry of the system. (In particular, what we can do is to double the $Z_2$ and $\t Z_2$ spins along the cut and cut through each pair.) Then on the boundary, each plaquette hosts a $Z_2$ variable $\tau_i$ and each vertex hosts a $\t Z_2$ variable $\si_i$.
It is then easy to see that on the 1D domain wall of $Z_2$ variables on the boundary is attached the 1D edge state of the $\t Z_2$ SPT phase. The $Z_2$ symmetry acts by flipping the plaquettes
\be
X \ket{\tau_i} = \ket{\bar{\tau_i}}
\ee
and the $\t Z_2$ symmetry acts on the $\si$ variables living on the $\tau$ domain walls as
\be
\t X \ket{\si_1,...\si_N}=\prod_{i=1}^N f(\si_i,\si_{i+1})\ket{\bar \si_1,...\bar \si_N}
\ee
where $f(\si_i,\si_{i+1})$ adds a phase factor of $i$ to each $\si$ domain wall and is $1$ if the neighboring $\si$'s are the same
\be
f(\si_i,\si_{i+1})=i^{\frac{1-\si_z^i\si_z^{i+1}}{2}}
\ee
as given in \Ref{Levin2012}.

\begin{figure}[htbp]
\begin{center}
\includegraphics[scale=0.3]{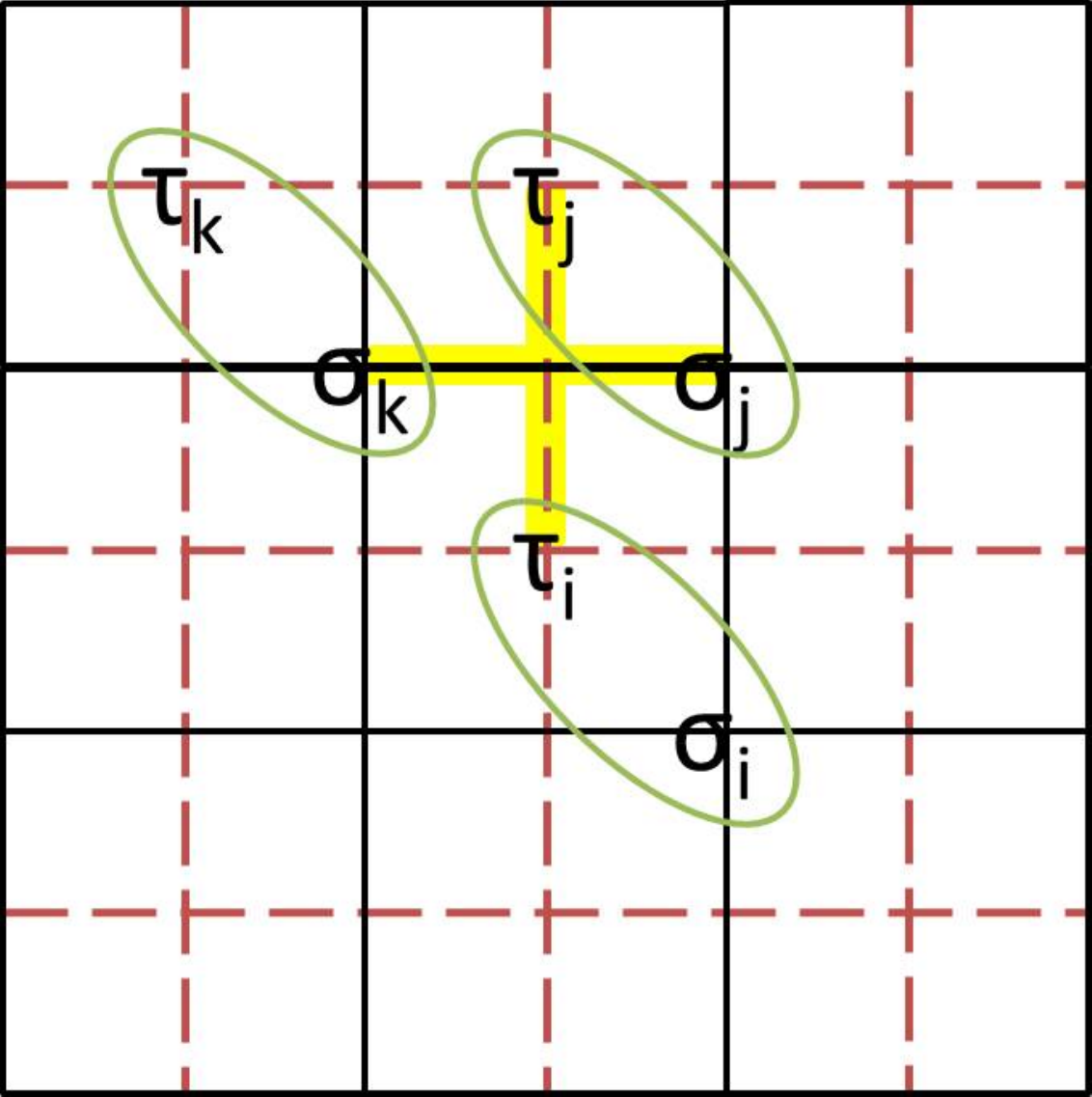}
\end{center}
\caption{Boundary of 3D SPT phase with $Z_2 \times \t Z_2$ symmetry. Each plaquette hosts a $Z_2$ variable $\tau$ and each vertex hosts a $\t Z_2$ variable $\si$. Circles represent local degrees of freedom on the 2D boundary labeled by group elements of $Z_2 \times \t Z_2$. Domain walls of $\tau$ live on solid lines and domain walls of $\si$ live on dashed lines.
}
\label{Z2_Z2_b}
\end{figure}

Equivalently, just like in the previous case, we can think of the two dimensional boundary as a system of independent degrees of freedom with one $Z_2$ variable $\tau$ per plaquette and one $\t Z_2$ variable $\sigma$ per vertex, as shown in Fig.\ref{Z2_Z2_b}. The $\tau$ domain walls lives on the solid lines and the $\sigma$ domain walls lines on the dashed lines. Then the $Z_2$ part of the symmetry simply acts by flipping the $Z_2$ variables $\tau_i$
\be
Z_2: \ \ \ket{\tau_i} \to \ket{\bar{\tau_i}}.
\ee
The $\t Z_2$ part of the symmetry action acts by flipping the $\t Z_2$ variables $\si_i$
\be
\t Z_2: \ \ \ket{\si_i} \to \ket{\bar{\si_i}}.
\ee
Moreover, the $\t Z_2$ symmetry also involves a phase factor $i$ whenever $\tau$ and $\si$ domain walls intersect. For example, at the shaded intersection shown in Fig.\ref{Z2_Z2_b}, the phase factor is given by
\be
i^{\frac{1-\si_z^j\si_z^{k}}{2}\frac{1-\tau_z^i\tau_z^{j}}{2}}
\ee

Such a symmetry action can be shown to be related to cocycles in a similar way as in section \ref{2D_GC}. In particular, in the group cohomology construction as reviewed in appendix \ref{app:GC_construct}, the boundary local degrees of freedom are labeled by group elements $\al_i$ of the symmetry group and the action of the symmetry operator involves two parts:
\begin{itemize}
\item{1. changing the local states $\ket{\al_i}$ to $\ket{\al\al_i}$,}
\item{2. multiplying a phase factor given by nontrivial cocycles to each triangle of $\al_i$, $\al_j$ $\al_k$.}
\end{itemize}
More specifically, on the 2D boundary of 3D SPT phases, the symmetry acts as
\be
O(\al)\ket{\al_1,....,\al_N}=\prod f^{\al}(\al_i,\al_j,\al_k)\ket{\al\al_1,...,\al\al_N}
\label{sym_2d}
\ee
where $f^{\al}(\al_i,\al_j,\al_k)$ is a phase factor given by the nontrivial 4th cocycle $\om_4$
\be
f^{\al}(\al_i,\al_j,\al_k)=\om_4(\al_i^{-1}\al_j,\al_j^{-1}\al_k,\al_k^{-1}\al^{-1},\al)
\ee

To put the above symmetry action of $Z_2\times \t Z_2$ into this form, we can relate each plaquette with one of its vertices (the lower right corner, for example) and think of the $Z_2$ and $\t Z_2$ variables in this pair as labeling group element $\al_i=(\tau_i,\si_i)$ in the symmetry group, as indicated by circles in Fig.\ref{Z2_Z2_b}. Then the symmetry action labeled by $\al={\tau,\si}$ flips the variables as
\be
\ket{\al_i=(\tau_i,\si_i)} \to \ket{\al\al_i=(\tau\tau_i,\si\si_i)}
\ee
Moreover, the phase factor is related to the triangle of $\al_i=(\tau_i,\si_i)$, $\al_j=(\tau_j,\si_j)$ and $\al_k=(\tau_k,\si_k)$ (although it depends trivially on $\tau_k$ and $\sigma_i$). Including the dependence of the phase factor on the symmetry action applied, we have
\be
\begin{array}{l}
f^{\al}(\al_i,\al_j,\al_k) = 1
\text{\ \ if $\si \in \al$ is nontrivial} \\ \nonumber
f^{\al}(\al_i,\al_j,\al_k) = i^{\frac{1-\si_z^j\si_z^{k}}{2}\frac{1-\tau_z^i\tau_z^{j}}{2}}
\text{\ \ if $\si \in \al$ is trivial}
\end{array}
\ee
Now we can reorganize the variables and check that $\om_4(\al_i^{-1}\al_j,\al_j^{-1}\al_k,\al_k^{-1}\al^{-1},\al)=f^{\al}(\al_i,\al_j,\al_k)$ is indeed a nontrivial four cocycle using the condition given by appendix \ref{GC}.

Therefore, as discussed in appendix \ref{app:GC_construct}, the state constructed in this way corresponds to a nontrivial SPT phase with $Z_2\times \t Z_2$ symmetry and the physical features include gapless states on $Z_2$ domain walls on the boundary.



\subsection{Effective field theory description}
\label{3D_FT}


We access this 3+1D  topological phase by directly constructing the 2+1 dimensional boundary. While the surface is a conventional 2+1D bosonic system except in the way symmetries are implemented, which prohibit a fully symmetric and gapped surface. We expand the physical symmetry $Z_2\times Z_2$ to $\left ( Z_2\times Z_2 \right )\times U(1)$. Eventually we will break the $U(1)$ to $Z_2$ and identify it with one of the $Z_2$ generators.

We begin with a bosonic field on the surface: $b_1 \sim e^{i\phi_1}$ that is charged only under the $U(1)$ symmetry. A possible SPT surface state is to spontaneously break this $U(1)$ symmetry by condensing this boson. Restoring the symmetry requires condensing vortices $\psi$ of this bosonic field. However, for this to be the surface of a topological phase, the vortex transformation law must forbid a  vortex condensate that preserves the remaining  $Z_2\times Z_2$ symmetry. This can be achieved if the transformation law is projective, which implies at least a two fold degeneracy of the vortex fields hence $\psi= ( \psi_+,\, \psi_- )$. Intuitively, one may consider the elements of the group $Z_2\times Z_2 =\left \{ 1,\, g_X,\, g_Y,\, g_Z \right \}$ as representing 180 degree rotations about the x, y, z axes. The projective representation is then just the spin 1/2 doublet, with the nontrivial generators represented by the Pauli matrices $g_a =i\sigma^a$, thus $ \psi \,\xrightarrow{g_a}\, i\sigma^a \psi $. Physical operators, which can be directly measured, do not realize symmetry in a projective fashion, however, the vortex is a non-local object and can hence transform projectively. Thus the effective Lagrangian for the vortices is:

\be
{\mathcal L}_{\rm surface} = \sum _\sigma  |(\partial_\mu -ia_\mu) \psi_\sigma|^2 + K (\partial_\mu a_\nu - \partial_\nu a_\mu)^2
\ee

where, by the usual duality, the vortex fields are coupled to a vector potential $a$ whose flux is the number density of $b_1$  bosons. Now, a vortex condensate will necessarily break symmetry since the gauge invariant combination $N^a = \psi^\dagger \sigma^a \psi$ will acquire an expectation value. However,  this transforms like a vector under rotations and will necessarily break the $Z_2\times Z_2$ symmetry. Thus the theory passes the basic test for the surface state of a topological phase. This continues to be true when we break down the U(1) symmetry to $Z_2$ and identify it with one of the existing $Z_2$ symmetries. That is, the boson $b_1$ now transforms the $g_a$. This gives us three choices, which can be labeled by $a=X,\,Y,\,Z$ which identifies the generator that leaves $b_1$ invariant. Thus the phase labeled X has $b_1\rightarrow -b_1$ under the generators $g_Y,\,g_Z$ but is left invariant under $g_X$. Even under this reduced symmetry the surface states cannot acquire a trivial gap since the condensate of $b_1$ continues to break some of the symmetries. Now, there are three distinct nontrivial phases implied by this construction labeled $a=X,\,Y,\,Z$ which, combined with the trivial phase  confirms the $Z_2\times Z_2$ classification of topological phases with this symmetry \cite{Chen2011}. With this reduced symmetry, additional terms can be added to the surface Lagrangian including $\Delta {\mathcal L}_X = -(\psi^\dagger \sigma^x \psi \cos \phi_1) $  We would like to now verify that domain walls at the surface carry protected modes along their length.

Now, consider breaking the symmetry at the surface down to a single $Z_2$.  When the remaining $Z_2$ generator is different from the one used to label the phase, we will see a protected mode is present along the surface domain walls. For example, in the setup here where the phase is labeled by $X$, consider breaking the symmetry down to just the $Z_2$ generated by $g_Z$. This is realized by condensing just $\psi_+$ or $\psi_-$ vortices.  Consider a domain wall in the x-y plane along $x=0$, where for $x>0$ ($x<0$) we have $\psi_+$ ($\psi_-$) condensed. The domain wall is the region of overlap of these condensates where we have  $\psi_+^\dagger \psi_- \sim e^{i\phi_2}$ which represents a vortex tunneling operator across the domain wall. The phase $\phi_2$ is therefore conjugate the boson phase $\phi_1$ as in a Luttinger liquid. Note however the transformation law under the remain $Z_2$ symmetry is: $\phi_{1,2}\overset{g_Z}\rightarrow \phi_{1,2} +\pi$, which is the transformation law of the protected edge of the 2D $Z_2$ SPT phase \cite{Levin2012,Lu2012a}. A similar conclusion can be drawn for surface domain walls, when the remaining $Z_2$ symmetry is $g_Y$. On the other hand, in this $X$ phase, preserving the $g_X$ symmetry does not lead to protected modes since $\phi_1$ is invariant under this generator and  can be locked at a particular value without breaking the symmetry.

Now, let us write down a 3+1D topological field theory that describes this phase. We will restrict to an abelian theory, which is not ideal given that the symmetry acts like spin rotation along orthogonal directions. Nevertheless the abelian theory gives us valuable insights.  Consider the two species of bosons $e^{i\phi_{1,2}}$ introduced earlier. In the phase labeled $X$ above,  the number density $n_1$ conjugate to $\phi_1$ is invariant under all transformations, while the number density $n_2$ changes sign under two of them.  We can always pick these to be for group elements $g_X$, $g_Y$ as above. Then we write down the following `BF+FF' theory:

\begin{equation}
2\pi {\mathcal L} = \epsilon (B_1\partial a_1+ B_2\partial a_2)+ \frac{\Theta}{2\pi} \epsilon \partial a_1 \partial a_2
\end{equation}

where $\epsilon$ is short for $\epsilon^{\mu\nu\lambda\sigma}$ and the indices are suppressed. Here, $2\pi  n_1 = \epsilon^{0ijk}\partial_iB_{1jk}$ is the number density of boson $b_1$, and $F_{1ij}= \partial_i  a_{1j}-\partial_j  a_{1i}$ represents the vortex lines of boson $b_1$. Thus the vector potentials $a_1$ are invariant under the transformations, but $a_2$ changes sign under $g_X,\,g_Y$. Therefore the coefficient of the second term in the equation above must be quantized $\Theta=0,\pi$ by the usual arguments. The latter case represents the topological phase. The unusual properties of domain walls
when the symmetry is broken down from $Z_2\times Z_2\rightarrow Z_2$, at the surface is readily deduced from this theory. A domain wall which breaks $g_X,\,g_Y$ occurs when $\Theta=\pi\rightarrow -\pi$ on the surface leads to the surface Luttinger liquid action:

\begin{equation}
{\mathcal S}_{LL}=\frac1{2\pi}\int \partial_x \phi_1 \partial_t\phi_2 \,dx\;dt
\end{equation}

which describes a domain wall located along $z=0,y=0$, where we have replaced $a_{1,2i}=\partial_i \phi_{1,2}$. These are just the phase fields we discussed above, and in particular, under the remaining $g_Z$ transformation they are both shifted by $\phi_{1,2}\overset{g_3}\rightarrow \phi_{1,2}+\pi$, which is the transformation property of the nontrivial edge of a $Z_2$ 2D SPT phase.

\section{Gauging the $Z_2$ Symmetry}
\label{Sec:gauging}
\begin{figure}[htbp]
\begin{center}
\includegraphics[scale=0.3]{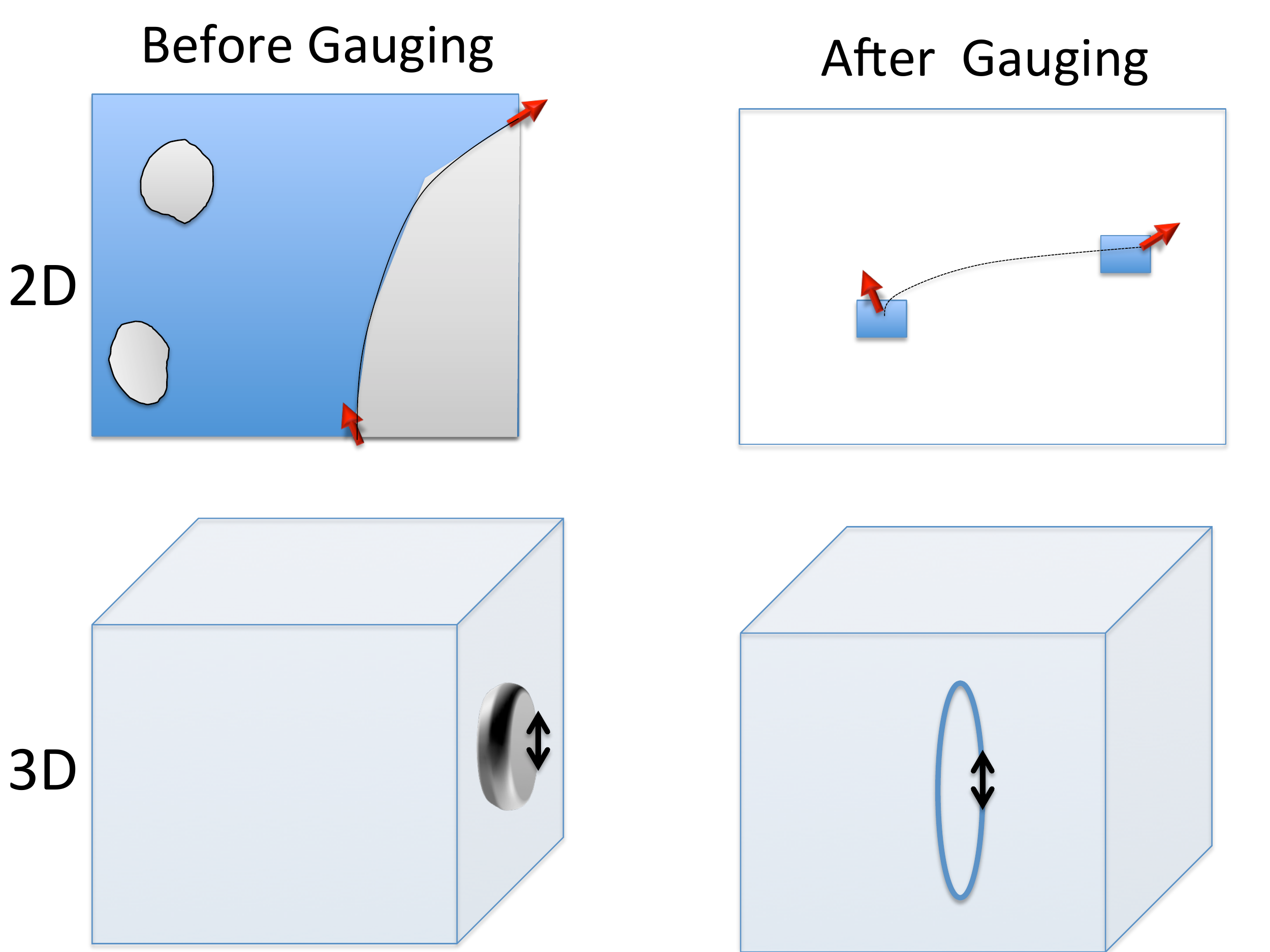}
\end{center}
\caption{Result of gauging the $Z_2$ symmetry. The gauge flux now allows the domain walls to terminate on them. They are point particles carrying a Kramers doublet in the $d=2$ system, and flux lines that carry a one dimensional edge state protected by the remaining $\t Z_2$ global symmetry in the $d=3$ system.
}
\label{gauging}
\end{figure}

In \Ref{Levin2012}, Levin and Gu discussed the topological phase in $d=2$ protected by just $Z_2$ symmetry. In that case, since there is no additional symmetry, we cannot interpret the domain walls with SPT phases in $d=1$. However, there is a different mechanism whereby an SPT phase is generated. The domain wall ends, which are now particles, carry fractional statistics (semionic statistics in this case). In fact this is readily seen from the symmetry transformation law for the edge state of this phase $\phi_{1,2}\rightarrow \phi_{1,2}+\pi$. Now, to generate a domain wall, both fields must be rotated, which is achieved by the domain wall operator $D(x) = e^{i(\phi_1(x)+\phi_2(x))/2}$. It is readily seen that $D(x)D(x')=i{\rm Sign}(x-x') D(x')D(x)$, which is the hallmark of semionic  statistics in 1D. One can now gauge the $Z_2$ symmetry, to obtain a topologically ordered state that is distinct from the regular $Z_2$ gauge theory\cite{Hung2012,Mesaros2012,Essin2012,Lu2013,Hung2013}. In fact the particle excitations in this theory are semions, and anti-semions, and this is termed the double-semion theory. Indeed,  the act of gauging the $Z_2$ symmetry simply implies the possibility of domain walls ending within the 2D sample. The location of these ends are just the gauge charges and fluxes. Given our experience with domain walls ending at the edge of the sample, where they have been shown to be simians, these end points are then expected to be semions.

We now apply this intuition to  the cased discussed here. First consider the $d=2$ SPT phase with  $Z_2\times Z_2^T$ symmetry. If the $Z_2$ part of the symmetry is gauged, domain walls could end within the sample. We now expect these ends, which are $Z_2$ fluxes, to carry a spin $1/2$ degrees of freedom. The gauged version this model is similar to the one discussed in \Ref{Yao2010}. From the cohomology classification we know that 2D phases with $Z_2\times Z_2^T$ symmetry form a $Z_2 \times Z_2$ group. Therefore, with this construction together with the nontrivial SPT phase which appears with just the $Z_2$ symmetry, we are able to account for all SPT phases possible in this system.

Now consider the $d=3$ SPT phase with $Z_2\times \t Z_2$ symmetry. Consider gauging $g_1$ the first $Z_2$ symmetry, but leaving the remaining $\t Z_2$ engaged so it continues to act as a global symmetry.  Now consider and inserting a gauge flux along a curve. It is readily seen that this flux curve will have an SPT edge state along it, specifically, the edge state of the $d=2$ SPT phase protected by $\t Z_2$ symmetry. It can therefore be distinguished from a trivial phase, where gauging one of the $Z_2$ degrees of freedom does not lead to protected excitations along  $Z_2$ flux lines.  To see this, note that gauging the symmetry just means that the domain walls of $g_1$ ends along the flux line. However, this domain wall is now in a $\t Z_2$ SPT phase. This follows from the wave function in Eqn. \ref{wfn3D}, since the loops on the surface bounded by this curve, formed by domains of $g_2$, are weighed to give a 2D Levin-Gu wave function\cite{Levin2012}. 

An interesting question is the result of gauging entirely the $Z_2\times \t Z_2$ symmetry. If we begin in the SPT phase this should lead to a $Z_2\times Z_2$ topological order that is distinct from the conventional one, which described by the deconfined phase of a $Z_2\times Z_2$ gauge theory. Characterizing these subtle differences in topological order is left to future work.

\section{1D SPT Phase with $Z_2\times Z_2$ symmetry: From Decorated Domain Walls to Model Hamiltonians}
\label{1Dz2z2}

\begin{figure}[htbp]
\begin{center}
\includegraphics[scale=0.3]{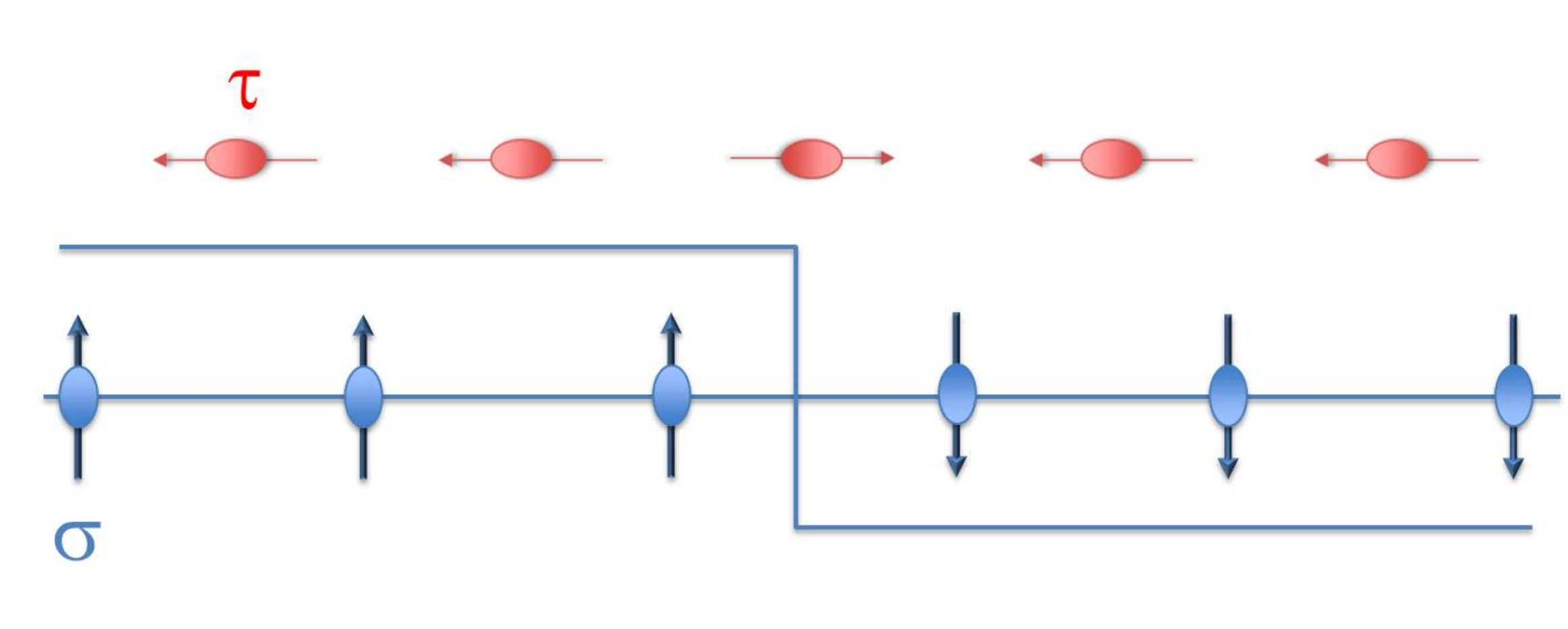}
\end{center}
\caption{An SPT phase in $d=1$ protected by $Z_2\times Z_2$ symmetry, that act on the two sets of spins $\sigma$, $\tau$. This phase emerges from the ordered phase of the $\sigma$ spins by condensing domain walls attached to spin flip excitations of the $\tau$ spins.
}
\label{Fig1D}
\end{figure}

We discuss how the well known Haldane phase in 1D can be understood within the framework of decorated domain walls. An advantage of this picture is that it leads directly to a model Hamiltonian that realizes this phase, and provides a simple rationale for the string order parameter of this state. Consider breaking down the full $SO(3)$ spin rotation symmetry to just $Z_2\times Z_2$ symmetry, which is sufficient to define this topological phase. 

The two $Z_2$s can be modeled by a pair of Ising models $\sigma,\,\tau$, and the ordered phases are given by $\langle \sigma^z\rangle \neq 0$ or $\langle \tau^z\rangle \neq 0$ or $\langle \sigma^z\tau^z\rangle \neq 0$. Consider beginning in the partially ordered phase $\langle \sigma^z\rangle \neq 0$ but $\langle \tau^z\rangle=0$. Now, the domain walls of $\sigma$ and the $Z_2$ quanta of $\tau$ are both gapped. If we condense the former, we enter the completely disordered phase. Condensing the latter leads to the completely ordered phase. But one can consider the following scenario. What happens when we condense the bound state of $\sigma$ domain wall and the $\tau$ $Z_2$ quanta (see figure \ref{Fig1D})?

It is readily seen that this describes the SPT phase. First, since we condense domain walls of $\sigma$, the $Z_2$ symmetry is restored. Note, although $\tau$ quanta are condensed, they are condensed along with the domain walls, so this is not a local operator, which implies that $Z_2$ is also unbroken.  So we have the full $Z_2 \times Z_2$ symmetry. The easiest way to see this is the SPT phase is to consider the `order' parameter for this phase, which is just the condensate $\rho = \tilde{\sigma}^z \tau^z$, where $\tilde{\sigma}^z$ is the disorder parameter that creates/destroys a domain wall in the $\sigma$ fields.  So we expect long range order in:

\begin{eqnarray*}
\langle \rho(i)\rho(j)\rangle &=& \langle\tau^z(i) \tilde{\sigma}^z(i)\tilde{\sigma}^z(j)\tau(j)\rangle \\
&=& \langle \tau^z(i) \left [\prod_{i<k<j}\sigma^x(k) \right ]\tau^z(j)\rangle
\end{eqnarray*}

However this is just the string order parameter for the $Z_2\times Z_2$ SPT phase\cite{Bahri2013}.

The following space-time picture motivates why it has edge states. Consider the path integral representation in imaginary time, with a  spatial boundary. Then space-time is a cylinder, the periodic direction being time. The condensate of Z$_2$ `charged' domain walls leads to world lines that sometimes intersect the boundary. When they do, since they carry Z$_2$ quantum number, they flip the spin at the edge - so there must be gapless edge degrees of freedom that fluctuate in time.

This picture also motivates the following Hamiltonian. First consider a pair of decoupled Ising Models, $\sigma$ and $\tau$ as shown in the figure. To enforce the binding of charge to domain walls we add the following Hamiltonian:

\begin{equation}
H_1= -\lambda \sum_i \left ( \sigma^z_i \sigma^z_{i+1}\tau^x_{i+1/2} +  \tau^z_{i-1/2} \tau^z_{i+1/2}   \sigma^x_i \right )
\end{equation}

The Hamiltonian $H_1^{\rm Cluster}$ \cite{Nielsen2006,Son2012} is just a set of commuting projectors, and has a unique ground state on a system with periodic boundary conditions. However, gapless edge states appear when the system is terminated at a boundary.


\section{3D SPT Phase with time reversal symmetry from decorated domain walls picture}
\label{3DT}

The decorating domain wall picture allows us to construct nontrivial SPT phases with time reversal symmetry also. In this section we are going to discuss 3D phases with $Z_2^T$ symmetry only, $Z_2^T\times U(1)$ symmetry and $Z_2^T\times SO(3)$ symmetry. In particular, the phase we construct with $Z_2^T$ symmetry only is beyond the cohomology classification.

Consider a 3D lattice with a spin $1/2$ sitting in each cube. Time reversal symmetry maps between spin up $\ket{\uparrow}$ to spin down $\ket{\downarrow}$ states together with the complex conjugation operation in this basis. 
Now we can decorate the domain wall in this spin configuration with some 2D states. First we need to specify the orientation of the domain walls as pointing from $\ket{\downarrow}$ to $\ket{\uparrow}$. Then we can put chiral states on the domain wall whose chirality matches the orientation of the domain walls. In particular, we put the Kitaev's $E_8$ state\cite{Kitaev2006} on the domain wall which is a bosonic 2D state with no fractional excitations in the bulk and a $c_-=8$ edge state. After this decoration, we sum over all possible spin configurations, together with the $E_8$ decoration. This wave function is invariant under time reversal symmetry which can be seen as follows. Time reversal maps between $\ket{\downarrow}$ and $\ket{\uparrow}$ changing the orientation not the potision of the domain walls. At the same time, time reversal maps the $E_8$ state to $-E_8$ state with inverse chirality. Therefore, the chirality of the $E_8$ is always consistent with the orientation of the domain wall and the total wave function is time reversal invariant. 

Such a construction gives rise to a 3D SPT state with only time reversal symmetry. Following similar arguments as in previous sections, we see that on the boundary of the system if we break time reversal symmetry in opposite ways on neighboring regions, then the domain wall between these regions carry a chiral edge state with $c_-=8$. This is half of what one can get in a pure 2D system without fractionalization. This is consistent with the field theory construction with $BF+FF$ term in Ref.\onlinecite{Vishwanath2012}. On the other hand, if time reversal symmetry is not broken, then the surface is gapless or has topological order. In fact, the surface topological state is that with three fermions\cite{Burnell2013} which cannot be realized in pure 2D system with time reversal symmetry.

Similar construction applies to 3D SPT phases with $Z_2^T\times U(1)$ and $Z_2^T\times SO(3)$ symmetry. In 2D with $U(1)$ and $SO(3)$ symmetry, there is an integer class of SPT phases with even integer quantized charge or spin Hall conductance.\cite{Lu2013,Chen2012,Liu2012,Senthil2012} Similar to the construction above, we can put the first nontrivial phase in this class to the domain wall of time reversal. The Hall conductance chirality should match the domain wall orientation. By summing over all spin configurations, we obtain a state with both time reversal and $U(1)$ or $SO(3)$ symmetry. One signature of the state would be even integer quantized Hall conductance on the time reversal domain wall on the 2D surface of the system.


\section{Relation to K\"{u}nneth formula for group cohomology of group $G=G_1\times G_2$}
\label{Kunneth}

If a group $G$ is the direct product of two subgroups $G=G_1\times G_2$, then the K\"{u}nneth formula for the group cohomology of $G$ can be written as\cite{Hung2012}
\be
\cH^{d+1}(G,U(1))=\sum_{k=0}^{d+1} \cH^k(G_1,\cH^{d+1-k}(G_2,U(1))) \label{KN}
\ee
which says that cohomology groups of $G$ in $d+1$ dimension can be obtained from cohomology groups of $G_1$ and $G_2$ in lower dimensions. Due to the relation between cohomology groups and SPT phases, this implies that SPT phases with $G$ symmetry in $d$ dimension can be constructed from SPT phases with $G_1$ and $G_2$ symmetry in lower dimensions.

When $k=0$, the term in the formula is
\be
\cH^0(G_1,\cH^{d+1}(G_2,U(1)))=\cH^{d+1}(G_2,U(1))
\ee
which means that some SPT phases with symmetry $G$ in $d$ dimension are just SPT phases with symmetry $G_2$ in $d$ dimension.

When $k=d+1$, the term in the formula is
\be
\cH^{d+1}(G_1,\cH^0(G_2,U(1)))=\cH^{d+1}(G_1,U(1))
\ee
which means that some SPT phases with symmetry $G$ in $d$ dimension are just SPT phases with symmetry $G_1$ in $d$ dimension.

Our domain wall construction correspond to the term with $k=1$
\be
\cH^1(G_1,\cH^d(G_2,U(1)))
\ee
$\cH^d(G_2,U(1))$ labels SPT phases with $G_2$ symmetry in $d-1$ dimensions. If $\cH^d(G_2,U(1))=M$ ($M=Z_n$, $Z$ for example), then the term becomes $\cH^1(G_1,M)$ which labels one dimensional representations of $G_1$ using $M$ coefficient. Therefore, this term says that some $d$ dimensional SPT phase with $G$ symmetry can be obtained from $d-1$ dimensional SPT phases with $G_2$ symmetry and one dimensional representation of $G_1$ with $M$ coefficient.

Our domain wall construction provides a physical interpretation of the above statement. With discrete $G_1$ group, as in the cases we are interested in, consider a $d$ dimensional configuration with group elements in $G_1$. The domain walls are then also labeled by group elements $g_1 \in G_1$ given by the difference of the group elements on the two sides of the wall. Then on the $d-1$ dimensional domain wall labeled by $g_1$, we can choose to put $d-1$ dimensional SPT phases from the class $M$. Our choice should be consistent with the fusion rules of the domain walls. That is, the phases $m^a$ and $m^b \in M$ which we choose to put onto domain walls $g^a_1$ and $g^b_1$ should combine into the phase $m^{ab}$ which we put onto the domain wall labeled by $g^a_1g^b_1$. In other words, this correspond to a one dimensional representation of group $G_1$ in coefficient $M$.

Therefore, the term with $k=1$ in the formula \ref{KN} correspond exactly to our domain wall construction. Note that this formula also gives the condition of when such domain wall construction can be consistent. In particular, when putting the $d-1$ dimensional SPTs onto the domain walls, we need to put them according to the corresponding one dimensional representation of $G_1$ in $M$. Suppose we have $M=Z_3$, then putting the first nontrivial one onto a $Z_2$ domain wall is not allowed because two $Z_2$ domain walls can merge to trivial and so should the corresponding SPTs.

Terms with higher $k$'s would correspond to constructing $d$ dimensional SPT phases with $G$ symmetry by putting $d-k$ dimensional SPT phases with $G_2$ symmetry onto $d-k$ dimensional defects in $G_1$ configurations.


\section{Conclusion and discussion}
\label{discussion}


In the previous sections we have presented the construction of a two dimensional SPT phase with $Z_2 \times Z_2^T$ symmetry by attaching Haldane chains, a one dimensional SPT phase with $Z_2^T$ symmetry, to the $Z_2$ domain walls in the 2D bulk and also a three dimensional SPT phase with $Z_2 \times \t Z_2$ symmetry by attaching the 2D SPT state with $\t Z_2$ symmetry to the domain wall of the $Z_2$ variables in the 3D bulk. Such a construction leads directly to the special topological feature of $Z_2$ domain walls on the boundary of the system: the $Z_2$ domain walls on the boundary carry gapless edge states of the other symmetry ($Z_2^T$ and $\t Z_2$). We established the nontrivial SPT order in the system by relating the symmetry action on the boundary of the system to nontrivial cocycles and also demonstrated how the SPT order can be properly described using field theories.

This domain wall construction also applies to  3D SPT phases with $Z_2^T$, $Z_2^T\times U(1)$ and $Z_2^T\times SO(3)$ symmetry, including the time reversal symmetric topological `superconductor' phase, which features chiral $E_8$ edge modes along surface domain walls that break time reversal symmetry. This state lies beyond the group cohomology classification.  The domain wall construction of attaching $d-1$ dimensional SPT states of $G_2$ symmetry to $d-1$ dimensional defects in $G_1$ configurations was shown to be related to one term in the K\"{u}nneth formula for the group cohomology of groups of the form $G_1\times G_2$. Other terms in the formula may be related to attaching lower dimensional SPT states with $G_2$ symmetry to lower dimensional defects in $G_1$ configurations, exploring which is left to future work. A 1D SPT phase with $Z_2\times Z_2$ symmetry, when approached in this manner, readily suggests a parent Hamiltonian. It would be interesting to find physically well motivated Hamiltonians that realize the higher dimensional topological phases as well. The physical viewpoint on SPT phases described in this work may help guide such a search.  



\acknowledgments

XC would like to thank Xiao-Gang Wen for pointing out the relation between the domain wall construction and the K\"{u}nneth formula. XC is supported by the Miller Institute for Basic Research in Science at UC Berkeley. AV thanks T. Senthil, Ying Ran, Ehud Altman, Yasaman Barhi, Lukasz Fidkowski and Michael Levin for insightful discussions, and is supported by NSF- DMR 0645691. Yuan-Ming Lu is supported by the Office of BES, Materials Sciences Division of the U.S. DOE under contract No. DE-AC02-05CH11231.


\appendix


\section{Projective representation and group cohomology}
\label{GC}

Matrices $u(g)$ form a projective representation of symmetry group $G$ if
\begin{align}
u(g_1)u(g_2)=\om(g_1,g_2)u(g_1g_2),\ \ \ \ \
g_1,g_2\in G.
\end{align}
Here $\om(g_1,g_2) \in U(1)$ and $\om(g_1,g_2) \neq 1$, which is called the
factor system of the projective representation. The factor system satisfies
\begin{align}
 \om^{s(g_1)}(g_2,g_3)\om(g_1,g_2g_3)&=
 \om(g_1,g_2)\om(g_1g_2,g_3),
 \label{2cocycle_om}
\end{align}
for all $g_1,g_2,g_3\in G$, where $s(g_1)=1$ if $g_1$ is unitary and $s(g_1)=-1$ if $g_1$ is anti-unitary.
If $\om(g_1,g_2)=1, \ \forall g_1,g_2$, this reduces to the usual linear representation of $G$.

A different choice of pre-factor for the representation matrices
$u'(g)= \bt(g) u(g)$ will lead to a different factor system
$\om'(g_1,g_2)$:
\begin{align}
\label{omom}
 \om'(g_1,g_2) =
\frac{\bt(g_1)\bt^{s(g_1)}(g_2)}{\bt(g_1g_2)}
 \om(g_1,g_2).
\end{align}
We regard $u'(g)$ and $u(g)$ that differ only by a pre-factor as equivalent
projective representations and the corresponding factor systems $\om'(g_1,g_2)$
and $\om(g_1,g_2)$ as belonging to the same class $\om$.

Suppose that we have one projective representation $u_1(g)$ with factor system
$\om_1(g_1,g_2)$ of class $\om_1$ and another $u_2(g)$ with factor system
$\om_2(g_1,g_2)$ of class $\om_2$, obviously $u_1(g)\otimes u_2(g)$ is a
projective presentation with factor system $\om_1(g_1,g_2)\om_2(g_1,g_2)$. The
corresponding class $\om$ can be written as a sum $\om_1+\om_2$. Under such an
addition rule, the equivalence classes of factor systems form an Abelian group,
which is called the second cohomology group of $G$ and is denoted as
$\cH^2[G,U(1)]$.  The identity element $1 \in \cH^2[G,U(1)]$ is the class that
corresponds to the linear representation of the group.

The above discussion on the factor system of a projective representation can be
generalized which gives rise to a cohomology theory of groups.

For a group $G$, let $M$ be a G-module, which is an abelian group (with
multiplication operation) on which $G$ acts compatibly with the multiplication
operation (\ie the abelian group structure) on M:
\begin{align}
\label{gm}
 g\cdot (ab)=(g\cdot a)(g\cdot b),\ \ \ \ g\in G,\ \ \ \ a,b\in M.
\end{align}
For the cases studied in this paper, $M$ is simply the $U(1)$ group and $a$ a
$U(1)$ phase. The multiplication operation $ab$ is the usual multiplication of
the $U(1)$ phases. The group action is trivial $g\cdot a=a$ ($g\in G$, $a\in
U(1)$) if $g$ is unitary and $g\cdot a = a^*$ if $g$ is anti-unitary.

Let $\om_n(g_1,...,g_n)$ be a function of $n$ group
elements whose value is in the G-module $M$. In other words, $\om_n:
G^n\to M$.  Let $\cC^n[G,M]=\{\om_n \}$ be the space of all such
functions.
Note that $\cC^n[G,M]$ is an Abelian group
under the function multiplication
$ \om''_n(g_1,...,g_n)= \om_n(g_1,...,g_n) \om'_n(g_1,...,g_n) $.
We define a map $d_n$ from $\cC^n[G,U(1)]$ to $\cC^{n+1}[G,U(1)]$:
\begin{align}
&\ \ \ \
(d_n \om_n) (g_1,...,g_{n+1})=
\nonumber\\
&
g_1\cdot \om_n (g_2,...,g_{n+1})
\om_n^{(-1)^{n+1}} (g_1,...,g_{n}) \times
\nonumber\\
&\ \ \ \ \
\prod_{i=1}^n
\om_n^{(-1)^i} (g_1,...,g_{i-1},g_ig_{i+1},g_{i+2},...g_{n+1})
\end{align}
Let
\begin{align}
 \cB^n[G,M]=\{ \om_n| \om_n=d_{n-1} \om_{n-1}|  \om_{n-1} \in \cC^{n-1}[G,M] \}
\end{align}
and
\begin{align}
 \cZ^n[G,M]=\{ \om_{n}|d_n \om_n=1,  \om_{n} \in \cC^{n}[G,M] \}
\end{align}
$\cB^n[G,M]$ and $\cZ^n[G,M]$ are also Abelian groups
which satisfy $\cB^n[G,M] \subset \cZ^n[G,M]$ where
$\cB^1[G,M]\equiv \{ 1\}$. $\cZ^n[G,M]$ is the group of $n$-cocycles and
$\cB^n[G,M]$ is the group of $n$-coboundaries.
The $n$th cohomology group of $G$ is defined as
\begin{align}
 \cH^n[G,M]= \cZ^n[G,M] /\cB^n[G,M]
\end{align}

When $n=1$, we find that $\om_1(g)$ satisfies
\be
\om_1(g_1)\om_1^{s(g_1)}(g_2)=\om_1(g_1g_2)
\ee
Therefore, the 1st cocycles of a group are the one dimensional representations of the group.

Moreover, we can check that the consistency and equivalence conditions (Eq. \ref{2cocycle_om} and \ref{omom}) of factor systems of projective representations are exactly the cocycle and coboundary conditions of 2nd cohomology group. Therefore, 2nd cocycles of a group are the factor systems of the projective representations of the group.

When $n=3$, from
\begin{align}
&\ \ \ \ (d_3 \om_3)(g_1,g_2,g_3,g_4)
\nonumber\\
&= \frac{ \om_3(g_2,g_3,g_4) \om_3(g_1,g_2g_3,g_4)\om_3(g_1,g_2,g_3) }
{\om_3(g_1g_2,g_3,g_4)\om_3(g_1,g_2,g_3g_4)}
\end{align}
we see that
\begin{align}
& \cZ^3[G,U(1)]=\{  \om_3|
\\
&\ \ \ \frac{ \om_3^{s(g_1)}(g_2,g_3,g_4) \om_3(g_1,g_2g_3,g_4)\om_3(g_1,g_2,g_3) }
{\om_3(g_1g_2,g_3,g_4)\om_3(g_1,g_2,g_3g_4)}
=1
 \} .
\nonumber
\end{align}
and
\begin{align}
& \cB^3[G,U(1)]=\{ \om_3| \om_3(g_1,g_2,g_3)=\frac{
\om_2^{s(g_1)}(g_2,g_3) \om_2(g_1,g_2g_3)}{\om_2(g_1g_2,g_3)\om_2(g_1,g_2)}
 \},
\label{3coboundary}
\end{align}
which give us the third cohomology group
$\cH^3[G,U(1)]=\cZ^3[G,U(1)]/\cB^3[G,U(1)]$.

\section{Group cohomology and symmetry action on the edge state of SPT phases}\label{app:GC_construct}

In establishing the nontrivial SPT order for the models discussed in section \ref{2D} and \ref{3D}, we used the connection between group cocycles and the symmetry action on the edge degrees of freedom. Here we review this connection established in \onlinecite{Chen2011} where the specific form of the edge symmetry action is given in terms of cocycles for the exactly solvable models. Because of such an nontrivial form of symmetry action, the edge state of nontrivial SPT phases cannot be trivially gapped without breaking symmetry. Note that in generic models, symmetry action on the edge degree of freedom can take more general form and it is not completely clear (in 3D and higher) how to extract the cocycle information from the symmetry action. However for the models discussed in section \ref{2D} and \ref{3D}, the edge symmetry action takes exactly the form given in \onlinecite{Chen2011} and therefore allows us to identify the cocycle and establish the nontrivial SPT order in a straight forward way.

1D SPT phases are classified by projective representations of the symmetry group $G$ with factor system $\om_2(g_1,g_2) \in \cH^2[G,U(1)]$. If the edge degree of freedom has basis states labeled by group elements $\ket{g}$, $g\in G$, then one possible form of the projective representation is
\be
U(g)\ket{g_1}=\om_2(g_1^{-1}g^{-1},g)\ket{gg_1}
\ee
That is, the symmetry action changes group element basis from $g_1$ to $gg_1$ and also adds a nontrivial phase factor given by $\om_2(g_1^{-1}g^{-1},g)$. It can be checked that $U(g)$ does form projective representation with factor system $\om_2(g_1,g_2)$.

2D SPT phases are classified by the third cocycles of the symmetry group $G$, $\om_3(g_1,g_2,g_3) \in \cH^3[G,U(1)]$. The edge of the system can be thought of as a local 1D system with a special symmetry action. If the degrees of freedom on the 1D edge have basis states labeled by group elements $\ket{g_i}$, $g_i\in G$, then one possible form of the effective symmetry action on the edge is
\be
\begin{array}{ll}
U(g)\ket{g_1,g_2...,g_N}=&\prod_i\om_3(g_i^{-1}g_{i+1},g_{i+1}^{-1}g^{-1},g) \times \\ \nonumber
&\ket{gg_1,gg_2,...,gg_N}
\end{array}
\ee
That is, the symmetry action changes group element basis from $g_i$ to $gg_i$ and also adds a nontrivial phase factor between every pair of nearest neighbor states given by $\om_3(g_i^{-1}g_{i+1},g_{i+1}^{-1}g^{-1},g)$.

3D SPT phases are classified by the fourth cocycles of the symmetry group $G$, $\om_4(g_1,g_2,g_3,g_4) \in \cH^4[G,U(1)]$. The edge of the system can be thought of as a local 2D system with a special symmetry action. To describe the symmetry action, we need to triangulate the surface such that the degrees of freedom live on the vertices of the triangulation. If the degrees of freedom have basis states labeled by group elements $\ket{g_i}$, $g_i\in G$, then one possible form of the effective symmetry action on the edge is
\begin{equation*}
\begin{array}{ll}
U(g)\ket{g_1,g_2...,g_N}=&\prod_{\Delta}\om_4(g_i^{-1}g_{j},g_j^{-1}g_k,g_k^{-1}g^{-1},g) \times \\
&\ket{gg_1,gg_2,...,gg_N}
\end{array}
\end{equation*}
That is, the symmetry action changes group element basis from $g_i$ to $gg_i$ and also adds a nontrivial phase factor on each triangle on the surface given by $\om_4(g_i^{-1}g_{j},g_j^{-1}g^k,g_k^{-1}g^{-1},g)$.

Symmetry action on the edge taking the particular forms given above is a sufficient but not necessary condition for the existence of SPT order. In general, the edge degrees of freedom do not have to have group element basis and the symmetry action can take more general form. However, for the discussion in this paper, these particular forms are enough to establish the SPT order in our models.

\section{Symmetry action on low energy states of Eq.\ref{XZ} from exact diagonalization} \label{ED}

In section \ref{2D_edge}, we studied one particular realization of the edge dynamics on the boundary of the two-dimensional SPT phase with $Z_2\times Z_2^T$ symmetry. The Hamiltonian governing the edge dynamics is given by
\be
\t H_e=\sum_i \tau_x^i\si_x^{i+1}+\tau_z^i\si_z^{i+1}+\si_x^i\tau_x^i+\si_z^i\tau_z^i
\label{XZ_app}
\ee
and the $Z_2$ symmetry acts on the edge as
\be
\t X=\prod_i \tau_z^i\si_z^i
\label{Z2_XZ_app}
\ee
while the $Z_2^T$ symmetry acts as
\be
\t T=\prod_i \Big(\frac{\tau_x^i+\tau_x^{i-1}}{2}\si_x^i+ \frac{\tau_x^i-\tau_x^{i-1}}{2}i\si_y^i\Big) K
\label{Z2T_XZ_a}
\ee
In this section, we show how to extract the low energy effective action of the symmetry on the edge state from exact diagonalization.

The Hamiltonian is an $XY$ model (written in $xz$ plane here) on a spin $1/2$ chain and the low energy effective theory is known to be described by the compactified free boson theory with Lagrangian
\be
2\pi\mathcal{L}_{edge}=\partial_t\phi_1\partial_x\phi_2-v\left[\left(\frac{\partial_x \phi_1}{2}\right)^2+\left(\partial_x\phi_2\right)^2\right]
\ee
The low energy states are labeled by quantum numbers $n_k \in Z$ and $\bar{n}_k \in Z$, $k=0,1,2...$, where $n_0$ and $\bar{n}_0$ label the total angular momentum and the winding number of the boson field respectively and $n_k$ and $\bar{n}_k$, $k>0$, label the left/right moving boson modes. If we normalize the ground state energy to be $0$ and the first excited state energy to be $1/4$, then the energy of each low energy state is given by
\be
E=\frac{n_0^2}{4}+\bar{n}_0^2+\sum_{k>0}k(n_k+\bar{n}_k)
\ee
and the lattice momentum $p$ ($-\pi/a<p\leq \pi/a$) of each state is given by
\be
p=\frac{\pi}{a}\left[(-)^{\bar{n}_0}+\sum_{k>0}\frac{2k}{L}(n_k-\bar{n}_k)\right]
\ee
where $L$ is the total system size.
$n_0$ is given by the conserved $U(1)$ quantum number
\be
S_y=\sum_i(\si_y^i+\tau_y^i)
\ee

The boson field $\phi=\frac{\phi_1+\phi_2}{2}$ can be thought of as describing the direction the spin $1/2$ is pointing to in the $xz$ plane. Therefore, it corresponds to the spin state $\cos{\phi}\ket{\uparrow}+\sin{\phi}\ket{\downarrow}$. From this, it is easy to see that the $Z_2$ symmetry maps $\phi$ to $-\phi$. However, it is not easy to see the action of the time reversal symmetry on the low energy state because of its complicated form. In order to obtain this information, we perform exact diagonalization of the Hamiltonian, identify the $n_k$, $\bar{n}_k$ quantum numbers of the eigenstates from quantities like $E$, $p$, $S_y$ and find out how time reversal symmetry acts on these states.

Energy levels in the spectrum can be degenerate and $p$ and $S_y$ allows us to partially split the degeneracy. However, states with $\{n_0=0,\bar{n}_0,\{n_k,\bar{n}_k\}\}$ and $\{n_0=0,-\bar{n}_0,\{n_k,\bar{n}_k\}\}$ have the same $E$, $p$, and $S_y$. In order to tell them apart, we need to use our knowledge of the action of the $Z_2$ symmetry. Using the decomposition of the $\phi$ field
\be
\begin{array}{l}
\phi(x,t)=\phi_0+\frac{\pi_0vt}{L}+\frac{\bar{\pi}_0x}{L}+ \sum_{k>0} \frac{1}{\sqrt{4\pi k}}\\ \nonumber
\left[b_ke^{-ip(x-vt)}+b_k^{\dagger}e^{ip(x-vt)}+\bar{b}_ke^{ip(x+vt)}+\bar{b}_k^{\dagger}e^{-ip(x+vt)}\right]
\end{array}
\ee
where $\pi_0$ measures total angular momentum (with quantum number $n_0$), $\bar{\pi}_0$ measures winding number (with quantum number $\bar{n}_0$) and $b_k$, $\bar{b}_k$ are the annihilation operators for left/right moving boson modes (with occupation number labeled by $n_k$ and $\bar{n}_k$). Because the $Z_2$ symmetry maps $\phi$ to $-\phi$, it maps $\pi_0$, $\bar{\pi}_0$, $b_k(b_k^{\dagger})$, $\bar{b}_k(\bar{b}_k^{\dagger})$ all to minus themselves. Therefore, under $Z_2$ symmetry action, $\ket{n_0,\bar{n}_0,n_k,\bar{n}_k}$ goes to $(-)^{\sum_{k>0}n_k+\bar{n}_k}\ket{-n_0,-\bar{n}_0,\{n_k,\bar{n}_k\}}$. Therefore using the action of the $Z_2$ symmetry, we can further distinguish states with $\{n_0=0,\pm\bar{n}_0,\{n_k,\bar{n}_k\}\}$ and fix the relative phase between them. If we want to fix the global phase factor of these two states, we can calculate the action of complex conjugation on them, which is expected to act as $\ket{n_0,\bar{n}_0,\{n_k,\bar{n}_k\}}$ to $\ket{-n_0,\bar{n}_0,\{\bar{n}_k,n_k\}}$.

Now we are ready to look at each degenerate sector labeled by $E$, $p$, $S_y$ and see how symmetry acts on them. We discuss the first few sectors as an illustration of method. The spectrum is obtained by diagonalizing a system with 16 spin $1/2$'s.

First the ground state with $E=0$, $p=0$, $S_y=0$ is nondegenerate. If the phase factor is fixed such that the state is unchanged under complex conjugation, then it is invariant under both $Z_2$ and time reversal symmetry.

The first excited states with $E=1/4$ and $p=0$ are two-fold degenerate, one with $S_y=1$ and one with $S_y=-1$. The two states are $\ket{1,0,\{0,0\}}$ and $\ket{-1,0,\{0,0\}}$. If we fix gauge such that complex conjugation acts as $\begin{pmatrix} 0 & 1 \\ 1 & 0\end{pmatrix}$ in this subspace and $Z_2$ symmetry acts as $\begin{pmatrix} 0 & 1 \\ 1 & 0\end{pmatrix}$ in this subspace, then time reversal acts as $\begin{pmatrix} -1.0000-0.0065i & 0.0000-0.0000i \\ 0.0000-0.0000i & -1.0000-0.0095i\end{pmatrix}$.

The third energy level at $p=0$ is also two fold degenerate with states $\ket{2,0,\{0,0\}}$ and $\ket{-2,0,\{0,0\}}$. If we fix gauge such that complex conjugation acts as $\begin{pmatrix} 0 & 1 \\ 1 & 0\end{pmatrix}$ in this subspace and $Z_2$ symmetry acts as $\begin{pmatrix} 0 & 1 \\ 1 & 0\end{pmatrix}$ in this subspace, then time reversal acts as $\begin{pmatrix} 0.9619 - 0.0004i & 0.0381 - 0.0099i \\ 0.0381 - 0.0099i & 0.9619 - 0.0004i \end{pmatrix}$.

The third energy level also has two fold degeneracy at $p=1$ with states $\ket{0,1,\{0,0\}}$ and $\ket{0,-1,\{0,0\}}$. If we fix gauge such that complex conjugation acts as $\begin{pmatrix} 1 & 0 \\ 0 & 1\end{pmatrix}$ in this subspace and $Z_2$ symmetry acts as $\begin{pmatrix} 0 & 1 \\ 1 & 0\end{pmatrix}$ in this subspace, then time reversal acts as $\begin{pmatrix} -0.0380 + 0.6342i & -0.9621 + 0.0251i \\  -0.9621 + 0.0251i & -0.0380 + 0.6342i \end{pmatrix}$.

The third energy level also has two fold degeneracy at $p=\pm \frac{1}{L/2}$ with states $\ket{0,0,\{n_1=1,\text{others are 0}\}}$ and $\ket{0,0,\{\bar{n}_1=1,\text{others are 0}\}}$. If we fix gauge such that complex conjugation acts as $\begin{pmatrix} 0 & 1 \\ 1 & 0\end{pmatrix}$ in this subspace and $Z_2$ symmetry acts as $\begin{pmatrix} -1 & 0 \\ 0 & -1\end{pmatrix}$ in this subspace, then time reversal acts as $\begin{pmatrix} 0 & -1 \\ -1 & 0 \end{pmatrix}$.

Therefore, up to finite size inaccuracy, the action of time reversal on the low energy state is consistent with $\ket{n_0,\bar{n}_0,\{n_k,\bar{n}_k\}}$ to $(-)^{\sum_{k\geq 0} n_k+\bar{n}_k}\ket{n_0,-\bar{n}_0,\{n_k,\bar{n}_k\}}$.

\section{Chern-Simons theory description of 2D SPT phases}\label{app:2d}

In this section we briefly review the Chern-Simons approach to 2+1-D SPT phases introduced in \Ref{Lu2012a}. In particular the $U(1)\times U(1)$ Chern-Simons effective theory and effective edge theory of bosonic $Z_2\times Z_2^T$-SPT phases will be discussed.

It is believed that any Abelian gapped phases (all quasi particle excitations obey Abelian statistics) in 2+1-D can be described by $U(1)^N$ Chern-Simons theory\cite{Read1990,Wen1992,Frohlich1991}. In general the bulk effective theory for a 2+1-D gapped Abelian phase is
\bea\label{cs bulk theory}
&\mathcal{L}=\frac{\epsilon_{\mu\nu\lambda}}{4\pi}\sum_{I,J=1}^Na^I_\mu{\bf K}_{I,J}\partial_\nu a_\lambda^J+\cdots
\eea
where ${\bf K}$ is a real symmetric $N\times N$ matrix of integer arguments and $\cdots$ represents non-universal higher-order terms (such as Maxwell terms). The ground state degeneracy of the gapped phase is $|\det{\bf K}|^g$ on a Riemann surface of genus $g$. For bosonic SPT phases which only support bosonic excitations in their spectra (no aynons), we must require $\det{\bf K}=\pm1$. On the other hand, the gauge invariance of the bulk Chern-Simons theory (\ref{cs bulk theory}) on an open manifold suggests the existence of edge excitations on the manifold's boundary\cite{Wen1995}:
\bea
&4\pi\mathcal{L}_{edge}=\sum_{I,J}\partial_x\phi_I\Big({\bf K}_{I,J}\partial_t\phi_J-{\bf V}_{I,J}\partial_x\phi_J\Big)\label{cs edge theory}
\eea
where ${\bf V}$ is a positive-definite real symmetric matrix which depends on microscopic details of the system. $\{\phi_I\}$ are chiral boson fields which is quantized to have $2\pi$ periodicity: $\phi_I\simeq\phi_I+2\pi$. The chirality of edge modes is determined by the signature $(n_+,n_-)$ of matrix ${\bf K}$: each positive(negative) eigenvalue of ${\bf K}$ corresponds to a right(left)-mover on the edge. Hence for the edge states to be non-chiral with the same number of right- and left-movers, $n_+=n_-$ and $\text{dim}{\bf K}=N=$even are required. Besides, for a bosonic system the diagonal elements of matrix ${\bf K}$ are all \emph{even} integers while for a fermionic system, at least one diagonal element of ${\bf K}$ are \emph{odd} integers.

A key ingredient for the Chern-Simons approach to SPT phases is the implementation of symmetry in effective theory (\ref{cs bulk theory}) and (\ref{cs edge theory}). For a non-chiral gapped phase, in the absence of symmetry its edge states in general are always gapped due to backscattering. Such backscattering terms on the edge has the following generic form of
\bea
\Delta\mathcal{L}_{edge}=\sum_{\bf l}A_{\bf l}\cos(\sum_I{\bf l}_I\phi_I+C_{\bf l}).\label{cs edge backscatter}
\eea
where ${\bf l}=(l_1,l_2,\cdots,l_N)$ are integer vectors and $A_{\bf l},C_{\bf l}$ are all real parameters. These backscattering terms correspond to Higgs terms which condense bosonic quasi particles\cite{Note2} in the bulk\cite{Lu2012a}. In the presence of symmetry, however, certain backscattering terms (\ref{cs edge backscatter}) can be forbidden due to nontrivial symmetry transformations of chiral bosons $\{\phi_I\}$. In other words, symmetry can protect the edge states from backscattering, yielding gapless edge excitations. Since the universal properties of SPT phases are fully encoded in their boundary excitations, the fingerprints of a SPT phase is contained in the matrix ${\bf K}$ and how the edge chiral bosons transform under symmetry. In general for a symmetry group $G$, the chiral bosons transforms under symmetry operation $g\in G$ as
\bea
&g\phi_Ig^{-1}=\eta^gg\sum_J{\bf W}^g_{I,J}\phi_J+\delta\phi_I,\\
&\notag\big({\bf W}^g\big)^T{\bf K}{\bf W}^g=\eta^g{\bf K},~~~0\leq\delta\phi_I<2\pi.
\eea
where $\eta^g=\pm1$ for unitary (anti-unitary) symmetries and ${\bf W}^g\in GL(N,\mathbb{Z})$ is a $N\times N$ unimodular matrix. These symmetry transformations $\{{\bf W}^g,\delta\phi^g\}$ on the edge chiral bosons form a faithful representation of symmetry group $G$ for the 2+1-D SPT phase. And the full set of data $[{\bf K},\{{\bf W}^g,\delta\phi^g\}]$ characterizes a SPT phase with symmetry group $G$.

With the same ${\bf K}$ matrix, different symmetry transformations sometimes correspond to the same SPT phase. How to identify different SPT phases in terms of their data $[{\bf K},\{{\bf W}^g,\delta\phi^g\}]$? It turns out with preserved symmetry group $G$, different gapped phases always form an Abelian group. The identity element of this group is the trivial phase, whose edge excitations can be gapped out without breaking any symmetry. For example, the state whose edge chiral bosons don't transform under symmetry belongs to the trivial phase. The other elements of the group are different SPT phases. The natural Abelian product rule $\bigoplus$ of two phases $[{\bf K},\{{\bf W}^g,\delta\phi^g\}]$ and $[\t{\bf K},\{\t{\bf W}^g,\t\delta\phi^g\}]$ is to take their matrix direct sum:
\bea
&\notag[{\bf K},\{{\bf W}^g,\delta\phi^g\}]\bigoplus[\t{\bf K},\{\t{\bf W}^g,\t{\delta\phi^g}\}]\\
&=[{\bf K}\oplus\t{\bf K},\{{\bf W}^g\oplus\t{\bf W}^g,\delta\phi^g\oplus\t{\delta\phi^g}\}].
\eea
It's easy to figure out the inverse of a phase is
\bea
&\notag[{\bf K},\{{\bf W}^g,\delta\phi^g\}]^{-1}=[-{\bf K},\{{\bf W}^g,\delta\phi^g\}].
\eea
Therefore two data $[{\bf K},\{{\bf W}^g,\delta\phi^g\}]$ and $[\t{\bf K},\{\t{\bf W}^g,\t\delta\phi^g\}]$ will represent the same  phase \emph{if and only if}
\bea
&[{\bf K},\{{\bf W}^g,\delta\phi^g\}]\bigoplus[-\t{\bf K},\{\t{\bf W}^g,\t\delta\phi^g\}]=\text{trivial}.
\eea

In the following we illustrate the above general principles in an example: bosonic SPT phases with $Z_2\times Z_2^T$ symmetry. We denote the generator of $Z_2$ and time reversal symmetry as $X$ and $T$ respectively. For a bosonic SPT phase, the simplest ${\bf K}$ matrix is the following $2\times2$ one
\bea
{\bf K}=\begin{pmatrix}0&1\\1&0\end{pmatrix}.
\eea
It turns out this $2\times2$ ${\bf K}$ is enough to describe all different bosonic SPT phases with $Z_2\times Z_2^T$ symmetry, which form an Abelian $Z_2\times Z_2$ group. These SPT phases transform in the following way under $Z_2\times Z_2^T$ symmetry\cite{Lu2012a}:
\bea\notag
&X\begin{pmatrix}\phi_1\\ \phi_2\end{pmatrix}X^{-1}=\begin{pmatrix}\phi_1+\pi\\ \phi_2+n_X\pi\end{pmatrix},\\
&T\begin{pmatrix}\phi_1\\ \phi_2\end{pmatrix}T^{-1}=\begin{pmatrix}-\phi_1\\ \phi_2+n_T\pi\end{pmatrix}.\notag
\eea
where $n_X=0,1$ and $n_T=0,1$ correspond to the 4 different $Z_2\times Z_2^T$-symmetric gapped phases. Let's label these 4 phases as $[n_X,n_T]$. Among them $[0,0]$ is the trivial phase, whose edge states can be gapped out by backscattering term $A\cos(\phi_2+C)$. All the other 3 phases are non-trivial SPT phases, whose edge modes cannot be gapped out without breaking $Z_2\times Z_2^T$ symmetry. However, it's straightforward to show that
\bea
&[n_X,n_T]\bigoplus[n_X,n_T]=trivial.\notag
\eea
and hence these 4 phases form a $Z_2\times Z_2$ group.

In the following we prove that the phase $[0,1]$ with symmetry transformations (\ref{sym2:new rep}) has a different but equivalent representation (\ref{sym2_Z2_phi})-(\ref{sym2_T_phi}). Consider a $[0,1]$ phase with edge chiral bosons $\{\phi_{1,2}\}$ transforming as (\ref{sym2:new rep}), is put together with a phase $\t\Phi$ whose edge chiral bosons $\{\t\phi_{1,2}\}$ transform as (\ref{sym2_Z2_phi})-(\ref{sym2_T_phi}). It's easy to check that the following backscattering terms which preserves $Z_2\times Z_2^T$ symmetry can be added to the edge:
\bea
&\notag A_1\sin(\phi_1+\t\phi_1),\\
&\notag A_1\sin(\t\phi_1-\phi_1),\\
&\notag A_2\cos\phi_2\cos\t\phi_2.
\eea
The first two terms localize the value of chiral boson fields $\phi_1$ and $\t\phi_1$ up to $\pi$, while the last term fixes this ambiguity. When these three backscattering terms all become relevant, the edge states will be fully gapped out without breaking any symmetry. This means $[0,1]\bigoplus\t\Phi=[0,1]\bigoplus[0,1]=$trivial and hence $\t\Phi=[0,1]$. And we've shown that (\ref{sym2_Z2_phi})-(\ref{sym2_T_phi}) is just a different but equivalent representation of the SPT edge with symmetry transformations (\ref{sym2:new rep}).

\section{Field Theory for the Edge Lattice Model of a 2D SPT Phase}
In this section we derive the low-energy long-wavelength effective theory for the edge states of Hamiltonian (\ref{1d edge ham}). Since this model describes the 1+1-D edge states of a gapped 2+1-D state, periodic boundary condition is always imposed. Therefore the usual duality transformation\cite{Kogut1979} needs to be modified, by introducing an extra $Z_2$ gauge field\cite{Levin2012} $\mu^z$. To be specific, we perform the following duality transformation to the spin-$1/2$ variables ${\vec\tau}^i$ and ${\vec\sigma}^i$ in 1+1-D Hamiltonian (\ref{1d edge ham}):
\bea
&\tau^i_x=\mu^z_{2i-1,2i}\lambda^x_{2i-1}\lambda^x_{2i},~~\tau_z^i=\mu^x_{2i-1,2i},\notag\\
&\sigma^{i+1}_x=\mu^z_{2i,2i+1}\lambda^x_{2i}\lambda^x_{2i+1},~~\sigma_z^{i+1}=\mu^x_{2i,2i+1},\notag\\
&\tau_z^i\sigma_z^i=\lambda^z_{2i-1},~~\tau_z^i\sigma_z^{i+1}=-\lambda^z_{2i}.
\eea
And the gauge-invariant constraints are
\bea
&\lambda^z_{2n-1}\mu^x_{2n-2,2n-1}\mu^x_{2n-1,2n}=1,\notag\\
&\lambda^z_{2n}\mu^x_{2n-1,2n}\mu^x_{2n,2n+1}=-1.\notag
\eea
It's straightforward to see that $\prod_{i}(-1)^i\lambda^z_i=1$. Since there are $L$ (length of the 1+1-d system) independent constraints on dual variables ${\vec\mu}_{i-1,i}$ and ${\vec\lambda}_i$, these dual variables faithfully represent the $2^L$-dimensional Hilbert space of original spin-$1/2$ variables ${\vec\tau}^i$ and ${\vec\sigma}^i$. And the Hamiltonian (\ref{1d edge ham}) can be rewritten by dual variables as
\bea
H_e=\sum_n\mu^z_{n-1,n}(\lambda_{n-1}^x\lambda_n^x+\lambda_{n-1}^y\lambda_n^y).
\eea
The low-energy effective Lagrangian of such a XY model can be easily obtained\cite{Sachdev2011B,Levin2012} as
\bea\label{edge lagrangian}
2\pi\mathcal{L}_{edge}=\partial_t\phi_1\partial_x\phi_2-{v}\Big[K(\frac{\partial_x\phi_1}2)^2+\frac1K(\partial_x\phi_2)^2\Big].
\eea
where Luttinger parameter $K=1$ in model (\ref{1d edge ham}) and $v=4a$ is the velocity of edge excitations ($a$ denotes the lattice constant). In the low-energy the chiral boson fields $\{\phi_{1,2}\}$ are related to local spin operators by
\bea
&\notag\frac{\partial_x\phi_2(x)}{\pi}\sim\frac{\lambda^z_{2n-1}+\lambda^z_{2n}}{2a}\sim\frac{\tau^n_z(\sigma^n_z-\sigma^{n+1}_z)}{2a},\\
&e^{\imth\phi_2(x)}\sim\sigma^n_z+\imth\tau^n_z,\\
&\notag e^{\imth\phi_1(x)}\sim\mu^z_{2n-1,2n}\lambda_{2n-1}^+\lambda_{2n}^+\sim\mu^z_{2n,2n+1}\lambda_{2n}^+\lambda_{2n+1}^+\\
&\notag\sim\tau^n_x(1-\sigma_z^n\sigma^{n+1}_z)+\imth\tau^n_y(\sigma^n_z-\sigma^{n+1}_{z})\\
&\notag\sim\sigma^{n+1}_x(\tau^{n+1}_x-\tau^n_x)+\imth\sigma_y^{n+1}(1-\tau^n_x\tau^{n+1}_x).
\eea
From these low-energy operator correspondence and symmetry operation (\ref{sym2_Z2})-(\ref{sym2_T}), it's clear the chiral bosons transform as
\bea\label{sym2_Z2_phi}
X\begin{pmatrix}\phi_1\\ \phi_2\end{pmatrix}X^{-1}=-\begin{pmatrix}\phi_1\\ \phi_2\end{pmatrix},
\eea
under $Z_2$ operation $X$ and
\bea\label{sym2_T_phi}
T\begin{pmatrix}\phi_1\\ \phi_2\end{pmatrix}T^{-1}=\begin{pmatrix}-\phi_1+\pi\\ \phi_2+\pi\end{pmatrix}
\eea
under time reversal operation $T$. As shown in Appendix \ref{app:2d}, such a symmetry transformation for the 1+1-D edge state of a $2+1$-D $Z_2\times Z_2^T$-SPT phase has another equivalent but different representation:
\begin{eqnarray}
&\notag X\begin{pmatrix}\phi_1\\ \phi_2\end{pmatrix}X^{-1}=\begin{pmatrix}\phi_1+\pi\\ \phi_2\end{pmatrix},\\
&T\begin{pmatrix}\phi_1\\ \phi_2\end{pmatrix}T^{-1}=\begin{pmatrix}-\phi_1\\ \phi_2+\pi\end{pmatrix}.\label{sym2:new rep}
\end{eqnarray}
In the new representation (\ref{sym2:new rep}) it's easy to figure out the following operator $\hat{D}$ creates a $Z_2$ domain wall:
\bea
\hat{D}(x)\sim e^{\imth\phi_2(x)/2}.\label{2d:z2 domain wall operator}
\eea
since we have the following commutator
\bea
\notag[\phi_1(x),e^{\imth\nu\phi_2(y)}]=\nu\pi\cdot\text{sgn}(y-x)\cdot e^{\imth\nu\phi_2(y)},~~~\forall~\nu\in\mathbb{R}.
\eea
due to the Kac-Moody algebra $[\phi_1(x),\partial_y\phi_2(y)]=2\pi\imth\delta(x-y)$ from Lagrangian (\ref{edge lagrangian}). It's straightforward to show that
\bea
T\hat{D}(x)T^{-1}=-\imth\hat{D}^\dagger(x),~~~T\hat{D}^\dagger(x)T^{-1}=\imth\hat{D}(x).\notag
\eea
Hence when time reversal acting twice the domain wall operator obtains a minus sign: $\hat{D}(x)\overset{T^2}\longrightarrow-\hat{D}(x)$. Therefore the low-energy effective theory of the edge states also predicts that a $Z_2$ domain wall on the edge of $Z_2\times Z_2^T$-SPT phase should carry the projective representation of time reversal symmetry, just as implied in the lattice model (\ref{1d edge ham}) with symmetry (\ref{sym2_Z2})-(\ref{sym2_T}).



\end{document}